\documentclass[journal]{IEEEtran}
\ifCLASSINFOpdf
\else
\fi
\usepackage{url}
\usepackage[hidelinks]{hyperref}
\usepackage[utf8]{inputenc}
\usepackage{graphicx}
\usepackage{amsmath}
\usepackage{amsfonts}
\usepackage{booktabs}
\usepackage{threeparttable}
\usepackage{algorithm}
\usepackage{algpseudocode}
\usepackage{graphics}
\usepackage{epsfig}
\usepackage{float}
\usepackage[noadjust]{cite} 

\begin{document}
%
\title{Robust Dynamic Network Embedding via Ensembles}
%
%
%

\author{Chengbin~Hou, Guoji~Fu, Peng~Yang, Zheng~Hu, Shan~He,
        and~Ke~Tang
\thanks{C. Hou, G. Fu, P. Yang, and K. Tang are with Department of Computer Science and Engineering, Southern University of Science and Technology, Shenzhen 518055, China. 
~~~C. Hou, and S. He are with School of Computer Science, University of Birmingham, Birmingham B15 2TT, United Kingdom.
~~~Z. Hu is with Huawei Technologies Co., Ltd., Shenzhen, China.
\protect\\ E-mail: chengbin.hou10@foxmail.com; fugj2017@student.sustech.edu.cn; yangp@sustech.edu.cn; hu.zheng@huawei.com; s.he@cs.bham.ac.uk; tangk3@sustech.edu.cn}
\thanks{© 2021 IEEE. Personal use of this material is permitted. Permission from IEEE must be obtained for all other uses.}
}

%
%

\markboth{Journal of \LaTeX\ Class Files,~Vol.~XX, No.~XX, MM~YYYY}%
{Shell \MakeLowercase{\textit{et al.}}: Bare Demo of IEEEtran.cls for IEEE Journals}
%



\maketitle

\begin{abstract}
Dynamic Network Embedding (DNE) has recently attracted considerable attention due to the advantage of network embedding in various fields and the dynamic nature of many real-world networks. An input dynamic network to DNE is often assumed to have smooth changes over snapshots, which however would not hold for all real-world scenarios. It is natural to ask if existing DNE methods can perform well for an input dynamic network without smooth changes. 
To quantify it, an index called Degree of Changes (DoCs) is suggested so that the smaller DoCs indicates the smoother changes.
Our comparative study shows several DNE methods are not robust enough to different DoCs even if the corresponding input dynamic networks come from the same dataset, which would make these methods unreliable and hard to use for unknown real-world applications.    
To propose an effective and more robust DNE method, we follow the notion of ensembles where each base learner adopts an incremental Skip-Gram embedding model. To further boost the performance, a simple yet effective strategy is designed to enhance the diversity among base learners at each timestep by capturing different levels of local-global topology. Extensive experiments demonstrate the superior effectiveness and robustness of the proposed method compared to state-of-the-art DNE methods, as well as the benefits of special designs in the proposed method and its scalability.

\end{abstract}


%
\IEEEpeerreviewmaketitle

\section{Introduction}
%
%
%
%
\IEEEPARstart{N}{etwork} embedding (a.k.a. network or graph representation learning) has been widely applied to various fields such as social networks, biological networks, telecommunication networks, knowledge graphs, and drug discovery \cite{hamilton2017representation,cui2019survey,goyal2018graph,zhang2020network,wu2020comprehensive}. Most previous network embedding methods are designed for static networks, while the real-world networks are often dynamic by nature \cite{holme2015modern,latapy2018stream,Kazemi2020Representation}, namely dynamic networks in which edges and/or nodes may be added and/or deleted at each snapshot of a dynamic network. Due to the dynamic nature of many real-world networks and the advantage of network embedding in various fields, Dynamic Network Embedding (DNE) has recently attracted considerable attention. DNE aims to efficiently learn node embeddings for each current network snapshot at each timestep by preserving network topology using current and historical knowledge, so that the latest embeddings can facilitate various downstream tasks.

A dynamic network in terms of a series of snapshots (graph streams) is often assumed to have smooth changes \cite{zhu2016scalable,li2017attributed,zhu2018high,zhang2018timers,ma2020community,du2018dynamic,zhou2018dynamic,goyal2017dyngem}, which serves as the input to DNE. However, the assumption of smooth changes over snapshots would not hold for all real-world scenarios. It is natural to ask if existing DNE methods can perform well for an input dynamic network without smooth changes. To quantify the smoothness of changes, we suggest an index called Degree of Changes (DoCs)\footnote{The DoCs here is a global index to quantify the smoothness of changes of a dynamic network over all snapshots or timesteps.}, which describes the average number of streaming edges between consecutive snapshots spanning a dynamic network. Note that, DoCs might be a way to quantify and rank the smoothness of changes for different dynamic networks. And the smaller DoCs corresponds to the smoother changes.

The DoCs of an input dynamic network depends on not only the nature of a network (e.g., a citation network versus an email network) but also how we preprocess it (e.g., 10 streaming edges per timestep versus 100 streaming edges per timestep). The existing general-purpose DNE methods are usually benchmarked on several network datasets, each of which is preprocessed by one special slicing setting that leads to a dynamic network with a special DoCs. However, the dynamic network with the special DoCs might not really fit the real-world requirement (e.g., it may require a smaller DoCs to update embeddings more frequently, or a larger DoCs to reduce the updating frequency). It remains unclear whether DNE methods can still perform well for the dynamic network with a different DoCs due to a different slicing setting on the same dataset. This motivates us to investigate the robustness of DNE to different input dynamic networks with different DoCs, especially when they come from the same dataset\footnote{We can more fairly compare and rank the smoothness of changes of dynamic networks using DoCs if they are generated from the same dataset.}.

Although there have existed quite a few DNE works \cite{zhu2016scalable,zhu2018high,li2017attributed,yu2020node,chen2018scalable,goyal2017dyngem,goyal2020dyngraph2vec,ma2020community,yu2018netwalk,du2018dynamic,mahdavi2018dynnode2vec,peng2020dynamic,hou2020glodyne,Sajjad2019Efficient,zheng2019addgraph,pareja2020evolvegcn,xu2020inductive,gong2020exploring,zhou2018dynamic,singer2019node,zhang2018timers}, they have not considered the effect of different DoCs of an input dynamic network to DNE methods. As a result, these methods might not be robust enough to different DoCs even if the corresponding input dynamic networks come from the same dataset, which is shown in our comparative study. 
However, the robustness of DNE to different DoCs is a desirable characteristic, as this would improve the reliability and usability of the DNE method while applying it to unknown real-world applications. 
To this end, we aim to propose an effective and more robust (w.r.t. DoCs) DNE method. 



Concretely, the proposed method follows the notion of ensembles where the base learner adopts an incremental Skip-Gram neural embedding approach. Furthermore, a simple yet effective strategy is designed to enhance the diversity among base learners at each timestep by capturing different levels of local-global topology using random walks with different restart probabilities. 
\textit{Intuitively}, the diversity enhanced ensembles encourage the base learners to learn from more diverse perspectives, such that it is more likely to have a part of (not necessarily all) base learners producing good embeddings at each timestep. The diversity enhanced ensembles therefore provide a better redundancy design to handle uncertainties (e.g., requiring different DoCs for different tasks) in generating a dynamic network and its time-evolving topological changes over snapshots of a dynamic network. 

The contributions are as follows:
\begin{itemize}
\item An index called DoCs is suggested to quantify the smooth changes of a dynamic network. We then investigate the robustness of DNE to different input dynamic networks with different DoCs, especially when they are generated from the same dataset. The comparative study reveals that the existing DNE methods are not robust enough to DoCs, and might not always prefer an input dynamic network with the smaller DoCs, i.e., the smoother changes, which is often assumed in existing DNE works.
\item An effective and more robust DNE method is proposed via the ensembles of incremental Skip-Gram embedding models at each timestep. To further boost the performance, we also propose a simple yet effective strategy to enhance the diversity among base learners by capture different levels of local-global topology.
\item The comparative study also demonstrates the superior effectiveness and robustness of the proposed method compared to six state-of-the-art DNE methods. The ablation study and parameter sensitivity analysis confirm the benefits of special designs such as the ensemble design and the diversity enhancement strategy. The scalability test verifies our theoretical complexity analysis.
\end{itemize}

The rest of the paper is organized as follows. Section II reviews the literature of DNE methods and distinguishes the proposed method from them. Section III formulates the DNE problem, describes the DNE method in details, and analyzes the complexity of the DNE algorithm. Section IV presents the experimental settings including benchmark datasets, compared methods, and evaluation protocol. The experimental results of comparative study, ablation study, parameter sensitivity, and scalability test are discussed in Section V. Section VI concludes the paper and points out potential future works.



\begin{figure}[t]
    \centering
    \includegraphics[width=0.5\textwidth]{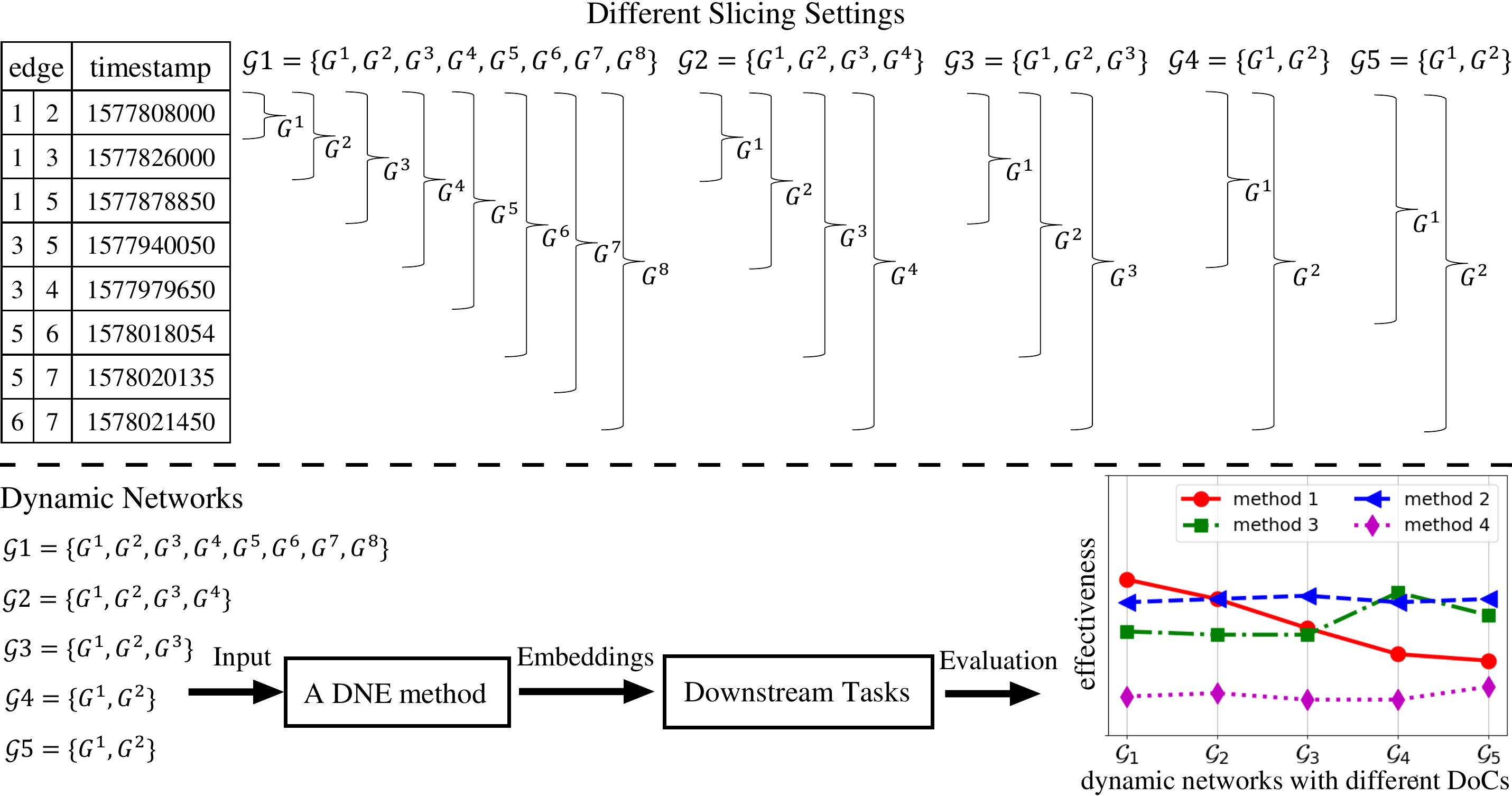}
    \caption{A toy example. The five input dynamic networks $\mathcal{G}1$,...,$\mathcal{G}5$ with different Degree of Changes (DoCs), constructed from the same dataset, are respectively fed to a DNE method and then evaluated by a downstream task. We could regard DoCs for $\mathcal{G}1$,...,$\mathcal{G}5$ as 1, 2, (3+2)/2=2.5, 4, 3 respectively. Despite the method 2 (in blue triangular) not always achieving the best, it may still be preferred  due to its good effectiveness and robustness.}
    \label{Fig1}
\end{figure}

\section{Literature Review}\label{literature}
Most existing DNE methods are developed based on Static Network Embedding (SNE) methods. For SNE methods, there have been several good surveys \cite{hamilton2017representation,cui2019survey,goyal2018graph,zhang2020network,wu2020comprehensive}. In general, SNE methods can be divided into four main categories. First, the \textit{Matrix Factorization (MF)} based approach \cite{cao2015grarep,ou2016asymmetric} encodes the desirable topological information into an n-by-n matrix, and then employs matrix factorization techniques to keep the top-d eigenvectors, which finally form the output embedding matrix. Second, the \textit{Auto-Encoder (AE)} based approach \cite{wang2016structural,cao2016deep} also encodes the desirable topological information into an n-by-n matrix, and then feeds each row of input matrix to an auto-encoder for training. The middle layer of the trained auto-encoder gives output embeddings. Third, the \textit{Skip-Gram (SG)} based approach \cite{perozzi2014deepwalk,grover2016node2vec} encodes the desirable topological information via a sampling strategy (e.g., random walkers) into node sequences. The node sequences can be treated as the network language, so that the Skip-Gram embedding model \cite{mikolov2013distributed} can be adopted for learning node embeddings. And fourth, the \textit{Graph Convolution (GC)} based approach \cite{kipf2016semi,hamilton2017inductive} defines node neighbors for each node by the first order proximity. For each node, node attributes are aggregated based on its neighbors so as to train a shared parametric mapping function. The trained function is finally used to infer node embeddings. Next, we review DNE based on the categories and methodologies of SNE.

\textit{MF based DNE.} The key challenge is how to update previous factorization results based on the difference between current and previous input matrices. \cite{zhu2018high,yu2020node} employ the matrix perturbation theory to resolve the challenge; \cite{li2017attributed} additionally considers the dynamic changes of node attributes; and \cite{zhang2018timers} discovers the accumulated error of such incrementally updating mechanism and discusses when to restart the matrix factorization from scratch. Unlike these works, \cite{zhu2016scalable} and \cite{chen2018scalable} iteratively minimize the reconstruction error of the current input matrix using previous factorization results (as a good initialization) for the faster convergence. 

\textit{AE based DNE.} One key challenge is how to fit the architecture of an Auto-Encoder for the current input matrix. When there is a new node leading to the increasing dimensionality of current input matrix, one neuron should be added to the first layer of the Auto-Encoder so as to make it runnable. \cite{goyal2017dyngem} proposes a heuristic strategy to adjust the Auto-Encoder to handle the challenge. \cite{ma2020community,goyal2020dyngraph2vec,yu2020node} also adopt an Auto-Encoder to incrementally learn node embeddings. But these works do not adjust the architecture of the Auto-Encoder and hence, they need to fix the node size in advance.

\textit{SG based DNE.} The key challenge is how to generate appropriate training samples to continuously train the Skip-Gram model. \cite{du2018dynamic,mahdavi2018dynnode2vec,peng2020dynamic,Sajjad2019Efficient} determine most affected nodes based on the changes between current and previous snapshots, and then employ a sampling strategy to generate the training samples related to the most affected nodes. \cite{hou2020glodyne} additionally considers inactive subnetworks, i.e., no change occurs for several timesteps, and generates training samples for the inactive subnetworks to better preserve the global topology.

\textit{GC based DNE.} Unlike above three categories, some SNE methods in this category, e.g., \cite{hamilton2017inductive} can be directly applied to DNE problem as they aim to learn a shared parametric mapping function, which can be treated as a stacked deep neural network where the number of neurons in the first layer is the dimensionality of node attributes. Therefore, there is no need to modify the neural network architecture even if there is a new node. It worth noticing that \cite{zheng2019addgraph,gong2020exploring,pareja2020evolvegcn,xu2020inductive} additionally employ the attention mechanism or Recurrent Neural Network (RNN) to capture the temporal dependency.

Moreover, some DNE methods might not be solely classified into one of the above four categories. For instance, \cite{singer2019node} applies a SNE method to obtain node embeddings at each timestep, and then feeds them sequentially to an RNN to capture the temporal dependency. \cite{zhou2018dynamic} considers the triadic closure process, social homophily, and temporal smoothness in the objective function, and then optimizes it based on the existing edges of each snapshot. 

Unlike the above DNE methods (for embedding a discrete-time dynamic network), \cite{nguyen2018continuous} takes the input of edge streams (for embedding a continuous-time dynamic network), and defines a temporal awareness random walk sampling strategy to generate training samples for continuously training the Skip-Gram model. Other works about continuous-time dynamic network embedding can be found in the survey \cite{Kazemi2020Representation}.

The proposed method, which takes snapshots as the input and Skip-Gram embedding approach as the base learner of ensembles, belongs to SG based DNE. But distinguished from existing works, this work aims to propose a more \textit{robust} DNE method w.r.t. DoCs, and introduces the \textit{ensembles} to DNE to improve its effectiveness and robustness.

\section{The Proposed Method}
\subsection{Notations and Definitions}
\textit{Definition 1: A Static Network.} $G(\mathcal{V},\mathcal{E})$ denotes a static network where $\mathcal{V}=\{v_1,...,v_n\}$ is a set of nodes; $\mathcal{V} \times \mathcal{V} \mapsto \mathcal{E}$ gives a set of edges; and $|\mathcal{V}|$ and $|\mathcal{E}|$ indicates the size of each set. The adjacency matrix is denoted as $\mathbf{A} \in \mathbb{R}^{|\mathcal{V}| \times |\mathcal{V}|}$ where $A_{ij}$ is the weight of edge $e_{ij}$; if $A_{ij} \neq 0$, the edge $e_{ij} \in \mathcal{E}$; and if $A_{ij} = 0$, there is no edge between $v_i$ and $v_j$. 

\textit{Definition 2: Static Network Embedding (SNE).} Given a static network $G(\mathcal{V},\mathcal{E})$, SNE aims to find a mapping $f: \mathcal{V} \mapsto \mathbf{Z}$ where $\mathbf{Z} \in \mathbb{R}^{|\mathcal{V}| \times d}$ is the output embedding matrix with a set of node embeddings; each row vector $\mathbf{Z}_i \in \mathbb{R}^d$ corresponds to node $v_i$ in $\mathcal{V}$; and $d \ll |\mathcal{V}|$ is the user-specified embedding dimensionality. The objective is to best preserve network topology while learning node embeddings.

\textit{Definition 3: A Dynamic Network.} $\mathcal{G}=\{G^0,G^1,...,G^t,...\}$ denotes the snapshots of a dynamic network at each timestep (i.e., graph streams). Each snapshot $G^t(\mathcal{V}^t,\mathcal{E}^t)$ is indeed a static network at timestep $t$. For a dynamic network, new nodes would come with new edges, and so this work only considers new edges but allows new nodes to occur. 

\textit{Definition 4: Degree of Changes (DoCs).} The DoCs is defined as the average number of streaming edges between consecutive snapshots spanning a dynamic network. Concretely, as the toy example shown in Figure \ref{Fig1}, the DoCs for $\mathcal{G}1$,...,$\mathcal{G}5$ is (1*7)/7=1, (2*3)/3=2, (3+2)/2=2.5, 4/1=4, and 3/1=3 respectively. Although other definitions could exist, the definition here might be the straightforward one.

\textit{Definition 5: Dynamic Network Embedding (DNE).} Given a dynamic network $\mathcal{G}=\{G^0,...,G^t\}$ with the latest snapshot $G^t(\mathcal{V}^t,\mathcal{E}^t)$, DNE aims to find a mapping $f^t: \mathcal{V}^t \mapsto \mathbf{Z}^t$ where $\mathbf{Z}^t \in \mathbb{R}^{|\mathcal{V}^t| \times d}$ is the output embedding matrix with a set of node embeddings at timestep $t$; each row vector $\mathbf{Z}_i^t \in \mathbb{R}^d$ corresponds to node $v_i$ in $\mathcal{V}^t$; and $d$ is the user-specified embedding dimensionality. The objective is to best preserve network topology and its dynamics into $\mathbf{Z}^t$, so that the latest node embeddings in $\mathbf{Z}^t$ can help various downstream tasks achieve good effectiveness. 

Besides the effectiveness of DNE for a certain DoCs of an input dynamic network, this work also considers its robustness (i.e., a small standard deviation or variance of effectiveness) across different DoCs especially when the corresponding input dynamic networks are generated from the same dataset, as the toy example illustrated in Figure \ref{Fig1}.

\subsection{Method Description}
\begin{figure*}[htbp]
    \centering
    \includegraphics[width=0.955\textwidth]{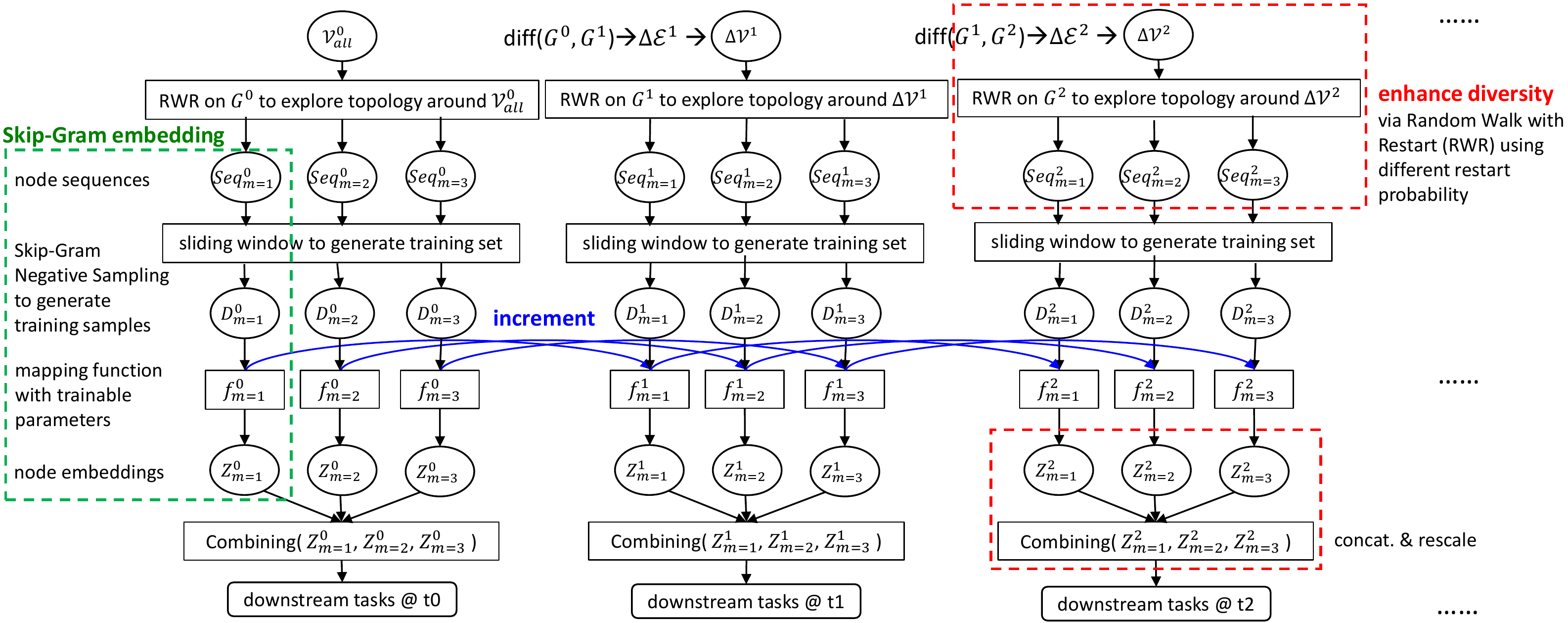}
    \caption{The overview of proposed method. At each timestep, multiple base learners (e.g., here are three) achieve ensembles. The base learner follows a Skip-Gram embedding approach. Between consecutive timesteps, each base learner inherits its previous base learner obeying an incremental learning paradigm. To enhance the diversity among base learners, random walk with restart using different restart probabilities is adopted to capture different levels of local-global topology (around the affected nodes by streaming edges). Embeddings from each base learner are concatenated and rescaled before downstream tasks.}
    \label{Fig2}
\end{figure*}

The proposed method is termed as \underline{S}kip-\underline{G}ram based \underline{E}nsembles \underline{D}ynamic \underline{N}etwork \underline{E}mbedding (SG-EDNE). It consists of four key components: Skip-Gram embedding approach, incremental learning paradigm, ensembles, and diversity enhancement as the overview illustrated in Figure \ref{Fig2}.

\subsubsection{Skip-Gram Embedding} \label{sec3.2.1}
A Skip-Gram embedding approach, an efficient neural network model \cite{shi2019mlne}, acts as the basic building block of the proposed method for learning node embeddings. At each current timestep $t$, the Skip-Gram Negative Sampling (SGNS) embedding model \cite{grover2016node2vec,hou2020glodyne} is adopted to learn node embeddings\footnote{The reasons of using Skip-Gram rather than other embedding approaches are as follows. While extending SNE to DNE approaches, first, the matrix factorization based and auto-encoder based approaches often require to set a fixed number of nodes in advance \cite{zhu2016scalable,zhu2018high,ma2020community,yu2018netwalk}, and hence are hard to handle unlimited new nodes. Second, the performance of graph convolution based approach \cite{kipf2016semi,hamilton2017inductive} largely depends on the node attributes, which is out of the consideration of this work. Third, the Skip-Gram based approach has been shown to be not only effective but also efficient to DNE \cite{hou2020glodyne,du2018dynamic,mahdavi2018dynnode2vec,peng2020dynamic,Sajjad2019Efficient}, and can easily handle node additions and deletions \cite{hou2020glodyne}.}, given the node sequences that capture desirable network topology (how to generate node sequences is presented in Section \ref{sec3.2.4}).

Specifically, a sliding window with length $s+1+s$ is employed to slide along each node sequence, so as to generate a series of positive node-pair samples by $(v_{center},v_{center+i})$, $i \in [-s,+s]$ and $i \neq 0$. 
In this way, the desirable network topology captured in node sequences (see Section \ref{sec3.2.4}) is now encoded into node co-occurrence $(v_c,v_i)$ statistics in $D^t$. We then maximize the log probability of node co-occurrence over all positive samples in $D^t$, i.e.,
\begin{equation}
    \max\sum\nolimits_{(v_c,v_i) \in D^t} {\log P(v_i \mid v_c)}
    \label{eq4}
\end{equation}
where $v_c$ is the center node, and $v_i$ is its nearby node with $1 \sim s$ orders proximity to $v_c$.

To define $P(v_i \mid v_c)$, it can be treated a binary classification problem \cite{levy2014neural}. Specifically, at timestep $t$, each positive sample $(v_c,v_i) \in D^t$ is distinguished from $q$ negative samples $(v_c,v_{i'})$s where $v_{i'}$ is drawn from a unigram distribution $P_{D^t}$ \cite{levy2014neural}. We then have $P(B = 1 \mid v_c,v_i) = \sigma {(\mathbf{Z}_c^t\cdot \mathbf{Z'}_i^t)}$ for observing a positive node-pair sample, and $P(B = 0 \mid v_c,v_{i'}) =1- \sigma {(\mathbf{Z}_c^t \cdot \mathbf{Z'}_{i'}^t)} = \sigma (-\mathbf{Z}_c^t \cdot \mathbf{Z'}_{i'}^t)$ for observing a negative node-pair sample. For all positive samples in $D^t$ and their corresponding $q$ negative samples, the objective is
\begin{equation}
    \begin{aligned}
    \max_{\mathbf{Z}^t,\mathbf{Z'}^t} \sum\nolimits_{(v_c,v_i) \in D^t} &\{\log\sigma (\mathbf{Z}^t_c\cdot \mathbf{Z'}^t_i) \\ 
    + \sum\nolimits_{i'=1}^q &\mathbb{E}_{v_{i'}\sim P_{D^t}}[\log\sigma (-\mathbf{Z}^t_c\cdot \mathbf{Z'}^t_{i'})] \}
    \end{aligned}
    \label{eq5}
\end{equation}
where $\sigma(x)=1/(1+\exp(-x))$; operator $\cdot$ denotes a dot product between vectors; and $\mathbf{Z}^t_c \in \mathbb{R}^d$ is the embedding vector from row $c$ of the trainable embedding matrix $\mathbf{Z}^t$ (between input and middle layers a.k.a. a project layer \cite{mikolov2013distributed}, and $\mathbf{Z}^t$ serves as DNE output after training), while $\mathbf{Z'}^t_i \in \mathbb{R}^d$ and $\mathbf{Z'}^t_{i'} \in \mathbb{R}^d$ are the embedding vector from column $i$ or $i'$ of another trainable matrix $\mathbf{Z'}^t$ (between middle and output layers) \cite{goldberg2014word2vec}. Eq. (\ref{eq5}) is optimized by stochastic gradient ascent over $D^t$ to learn these trainable embeddings. Intuitively, the more frequency two nodes co-occur in $D^t$, the closer or more similar their embeddings would be.

Note that, if the above approach only considers the snapshot at each timestep, i.e., treating as a static network embedding problem $\mathbf{Z}^t=f^t(G^t)$ at each timestep, we then need to retrain it from scratch at each timestep, which is time-consuming and cannot make use of historically learned knowledge.

\subsubsection{Incremental Learning} \label{sec3.2.2}
The incremental learning paradigm is employed to avoid retraining from scratch for better efficiency and exploit historically learned knowledge for better effectiveness. To further improve efficiency, we consider one step back learned knowledge in input to middle layer $\mathbf{Z}^{t-1}$ and in middle to output layer $\mathbf{Z'}^{t-1}$ which are indeed the weights of SGNS neural network model, one step back snapshot $G^{t-1}$, and current snapshot $G^t$. Formally, the incremental Skip-Gram based DNE is
\begin{equation}
    \mathbf{Z}^t = f^t(G^t,G^{t-1},\mathbf{Z}^{t-1},\mathbf{Z'}^{t-1})
    \label{eq6}
\end{equation}
where trainable embeddings $\mathbf{Z}^{t}$ and $\mathbf{Z'}^t$ in the neural network model $f^t$ are copied from its one step back $\mathbf{Z}^{t-1}$ and $\mathbf{Z'}^{t-1}$ in $f^{t-1}$ (or denoted as $f^t \gets f^{t-1}$) as the initialization before training. After training by Eq. (\ref{eq5}), we can take out the updated embedding matrix $\mathbf{Z}^t$ as the output of DNE at $t$.

It is worth noting that, Eq. (\ref{eq6}) holds for $t>0$ (online), and the latest two consecutive snapshots $G^{t}$ and $G^{t-1}$ provide the latest dynamics of streaming edges $\Delta\mathcal{E}^t$ (or given directly) which yields affected nodes $\Delta\mathcal{V}^t$ (see Figure \ref{Fig2}). When $t=0$ (offline), we have to train $\mathbf{Z}^0 = f^0(G^0)$ with randomized embeddings for all nodes in $\mathcal{V}^0$ of $G^0$ from scratch.

\subsubsection{Ensembles} \label{sec3.2.3}
The ensembles have been shown useful to incremental learning problems on Euclidean data for handling uncertainties to improve model effectiveness and robustness \cite{sun2018concept,Krawczyk2017Ensemble,losing2018incremental,gama2014survey}. Motivated by this, we attempt to employ ensembles to the DNE problem on graph (non-Euclidean) data.

At each timestep, we \textit{separately} train several base learners of the incremental Skip-Gram based DNE model in Eq. (\ref{eq6}) using the objective function in Eq. (\ref{eq5}) respectively. The objective function of the ensembles at each timestep $t$ becomes
\begin{equation}
    \begin{aligned}
    \sum\nolimits_{m=1}^{M} \{~\max_{\mathbf{Z}_m^t,\mathbf{Z'}_m^t} ~~ \sum\nolimits_{(v_c,v_i) \in D^t_m} &[~\log\sigma (\mathbf{Z}^t_{m,c}\cdot \mathbf{Z'}^t_{m,i}) \\
    + \sum\nolimits_{i'=1}^q \mathbb{E}_{v_{i'}\sim P_{D^t_m}}[\log\sigma &(-\mathbf{Z}^t_{m,c}\cdot \mathbf{Z'}^t_{m,i'})]~]~\}
    \end{aligned}
    \label{eq7}
\end{equation}
where the number of models to train is $M$; $D^t_m$ is the training samples for base model $m$; and the output embedding matrix for base model $m$ is $\mathbf{Z}_m^t=f_m^t(G^t,G^{t-1},\mathbf{Z}_m^{t-1},\mathbf{Z'}_m^{t-1})$. Please refer to Eq. (\ref{eq5}) and Eq. (\ref{eq6}) for other notations.

After training, the embeddings from each base learner are concatenated to form the unified embeddings, i.e.,
\begin{equation}
    \mathbf{Z}^t = {\mathbf{Z}^t_{m=1} \oplus ... \oplus \mathbf{Z}^t_{m=M}}
    \label{eq8}
\end{equation}
where $\oplus$ denotes the concatenation operator\footnote{The output from each base learner is embeddings in our (unsupervised learning) problem rather than a score in most supervised learning problems. Hence, the combining strategy of the outputs of embeddings is different from usual strategies, e.g., weighted voting of scores. Apart from concatenation, we also tested element-wise mean, max, min, and sum over each dimension of embeddings. The concatenation operator often obtained superior embeddings (evaluated by downstream tasks) for $d=128$ by convention \cite{grover2016node2vec,perozzi2014deepwalk,hou2020glodyne}.}.
Note that, we keep the dimensionality of unified embeddings to be $d$, i.e., $\mathbf{Z}^t \in \mathbb{R}^{|\mathcal{V}^t| \times d}$ regardless of the number of base learners $M$ (but $d\geq M$). To achieve that, the dimensionality for each $\mathbf{Z}_m^t$ is $\mathbb{R}^{|\mathcal{V}^t| \times d_m}$ where $d_m$ takes the integer of $d/M$ and the reminder is added to $d_m$ if $m=M$.

\subsubsection{Diversity Enhancement} \label{sec3.2.4}
The diversity among base learners is recognized as a key factor to improve the performance of ensembles \cite{brown2005diversity,Krawczyk2017Ensemble,sun2018concept}. It is thus natural to enhance diversity for further boosting the effectiveness and robustness of the ensembles mentioned in Eq. (\ref{eq7}). 

To be more specific, we propose to enhance the diversity among node sequences (and therefore training samples $D_m^t$) before feeding to each incremental Skip-Gram based DNE model. The strategy to enhance the diversity is that, each base learner adopts a different restart probability of Random Walk with Restart (RWR) to explore a different level of local-global topology over snapshot $G^t$ starting from each node in $\Delta \mathcal{V}^t$ which is affected by streaming edges at timestep $t$.

For the restart probability $R_m$, a maximum $R^\text{max}$ is set to assign different restart probabilities to $M$ base learners by
\begin{equation}
    {R_{1}, R_{2},...,R_{M}}\ =\ {0,R^\text{max}/M,...,(M-1)R^\text{max}/M}
    \label{eq9}
\end{equation}
where $R^\text{max}$ and $M$ are two hyper-parameters. The reason of setting restart probabilities uniformly is that, this work does not intend to find the optimal restart probabilities for a specific application. But one may try nonuniform restart probabilities for a certain real-world application. Furthermore, we believe different restart probabilities can capture different levels of local-global topology (see Figure \ref{Fig8}). Considering two extremes: for $R_m=1$, RWR can only explore very limited local neighbors around a starting node; for $R_m=0$, RWR can collect more global information from a starting node. 

With Eq. (\ref{eq9}), a RWR using $R_m$ for model $m$ can be used to generate \textit{node sequences} at timestep $t$. Concretely, for each node in $\Delta\mathcal{V}^t$ (or $\mathcal{V}^0$ if $t=0$), $r$ truncated RWRs with length $l$ are conducted starting from it. For each RWR, the next node $v_j$ much jump back to the starting node (i.e., restart) if a random number $P_R$ from a uniform distribution $U[0,1]$ meets $P_R < R_m$. Otherwise, the next node $v_j$ is sampled based on the transition probability of its previous node $v_i$ given by
\begin{equation}
{P(v_j \mid v_i) } = {A_{ij}^t} \ / \ {{\sum\nolimits_{v_{j'} \in \mathcal{V}^t} A_{ij'}^t }}
\label{eq10}
\end{equation}
where $A_{ij}^t$ is the edge weight between node $v_i$ and node $v_j$ at timestep $t$, and there is no edge between them if $A_{ij}^t=0$.

Each base learner is trained separately to preserve a different level of local-global topology into embeddings. To keep the concatenated embeddings from $M$ base learners in the same scale, we conduct a rescaling operation over each column\footnote{Comparing to zero mean and unit variance scaling, we found [0,1] min-max scaling obtained better results in our cases, and is thus used in this work. The motivation of rescaling $\mathbf{Z}^t$ over each column is that we hope all $d$ hidden variables/features from $M$ base learners are fairly used in downstream tasks.} of $\mathbf{Z}^t$, which then serves as the input to downstream tasks. In this way, the diverse local-global topological information might be fairly preserved and used in downstream tasks.


\subsection{Algorithm and Complexity Analysis} \label{complexity}
The workflow of proposed method SG-EDNE during online stage (i.e., $t>0$) is summarized in Algorithm \ref{Alg1}. Note that, Section \ref{sec3.2.1} describes L6, L8; Section \ref{sec3.2.2} describes L1, L7; Section \ref{sec3.2.3} describes L2, L8, L9, L11; and Section \ref{sec3.2.4} describes L3, L5, L12. For offline stage (i.e., $t=0$), we still follow Algorithm \ref{Alg1} except that we have to train $\mathbf{Z}_m^0 = f_m^0(G^0)$ with random parameters for all nodes in $\mathcal{V}^0$ from scratch. One may also restart the algorithm after $t=0$ to reduce potential accumulated errors \cite{zhang2018timers}. For both online and offline stages, one can parallelize L4-L10, as there is no interaction among $M$ base learners.
\begin{algorithm}[htbp] 
\caption{SG-EDNE at timestep $t$ (online)} 
\label{Alg1} 
\begin{algorithmic}[1] 
\Require 
two latest snapshots of a dynamic network $G^{t-1},G^t$; previous trained mapping functions $f_{m=1}^{t-1},...,f_{m=M}^{t-1}$; number of base learners/models $M$; max restart probability $R^\text{max}$; walks per node $r$; walk length $l$; sliding window size $s$; negative samples per positive sample $q$; embedding dimensionality $d$
\Ensure embedding matrix $\mathbf{Z}^t \in \mathbb{R}^{|\mathcal{V}^t| \times d}$
\State find affected nodes $\Delta \mathcal{V}^t$ by new edges based on $G^{t-1},G^t$
\State assign dimensionality $d_m$ to base learner based on $d$, $M$
\State assign restart probability $R_m$ to base learner based on $R^\text{max}$ and $M$ by Eq. (\ref{eq9})
\For{base learner $m$=1 to $M$}
\State generate node sequences $Seq_m^t$ by RWR with $r,l,R_m$ starting from each node in affected nodes $\Delta \mathcal{V}^t$
\State generate training samples $D_m^t$ by SGNS with $s, q$
\State copy trainable parameters to $f_m^{t}$ from $f_m^{t-1}$
\State train $f_m^{t}$ using $D_m^t$ by Eq. (\ref{eq5})
\State take out $\mathbf{Z}_m^t$ from $f_m^t$
\EndFor 
\State combine $\mathbf{Z}_{m=1}^t$,...,$\mathbf{Z}_{m=M}^t$ to obtain $\mathbf{Z}^t$ by Eq. (\ref{eq8}) 
\State rescale $\mathbf{Z}^t$ over each column \\
\Return $f_{m=1}^{t},...,f_{m=M}^{t}$ and $\mathbf{Z}^t$
\end{algorithmic} 
\end{algorithm}

The ensembles of $M$ base learners does not increase time complexity, as $M$ is a small constant. Consequently, we ignore the effect of $M$ as well as other constants such as $r$, $l$, and $d$ in the following complexity analysis.
During the \textit{online} stage that follows the incremental learning paradigm as described in Algorithm \ref{Alg1}, L5 requires $O(|\Delta\mathcal{V}^t|)$, as the algorithm conducts RWR for $|\Delta\mathcal{V}^t|$ nodes using alias sampling method \cite{grover2016node2vec}. The complexity of L6-L9 is $O(|\Delta\mathcal{V}^t|)$, since the algorithm generates training samples and accordingly trains SGNS based on $O(|\Delta\mathcal{V}^t|)$ node sequences. Moreover, the complexity for L1, L11, and L12 is $O(|\mathcal{V}^t|)$ respectively, while the complexity for L2 and L3 is $O(1)$ respectively. In summary, the overall complexity for the \textit{online} stage is $O(|\Delta\mathcal{V}^t|+|\mathcal{V}^t|)$. While for the \textit{offline} stage, i.e., $t=0$ or in case of restarting algorithm, the overall complexity becomes $O(|\mathcal{V}^t|)$, since L5-L9 are now based on $O(|\mathcal{V}^t|)$ node sequences.

\section{Experimental Settings}
\subsection{Datasets} \label{Sec4.1}
In this work, the independent variable to control is DoCs. Consequently, we adopt five different slicing settings to each dataset to generate five dynamic networks $\mathcal{G}1,...,\mathcal{G}5$ with five different DoCs. The statistics are presented in Table \ref{Tab1}. Given the same dataset, the more slices often lead to the smaller DoCs of the corresponding generated dynamic network.

All datasets\footnote{DNC-Email (\url{http://networkrepository.com/email-dnc.php}) is the email network of Democratic National Committee email leak in 2016. College-Msg (\url{http://snap.stanford.edu/data/CollegeMsg.html}) is the messaging network of an online social network at the University of California, Irvine. Co-Author (\url{http://networkrepository.com/ca-cit-HepPh.php}) is the collaboration network of authors from the papers of high energy physics phenomenology in arXiv. FB-Wall (\url{http://networkrepository.com/ia-facebook-wall-wosn-dir.php}) is a social network for Facebook users interacting in their wall posts. Wiki-Talk (\url{http://snap.stanford.edu/data/wiki-Talk.html}) is the network of registered Wikipedia users editing in each other's talk pages for discussions.} \cite{Rossi2015dataset,Leskovec2014dataset} are originally in \textit{edge streams} and each edge has an Unix timestamp.
Because of a large volume of experiments and some DNE methods encountering the out of memory (OOM) issue, we take out partial edges up to date 19951231, 20070331, and 20050331 on Co-Author, FB-Wall, and Wiki-Talk respectively. For other datasets, their full datasets are used. For each dataset, five dynamic networks are generated in the following way: 1) order edge streams by timestamps; 2) take the first 1/5 edges to establish the initial snapshot $G^0$; 3) the rest of edges are roughly and evenly divided into 20, 40, 60, 80, and 100 slices respectively for dynamic networks $\mathcal{G}1$, $\mathcal{G}2$, $\mathcal{G}3$, $\mathcal{G}4$, and $\mathcal{G}5$; and 4) the largest connected component, unweighted, and undirected $G^1$ of $\mathcal{G}1$ is given by $G^0$ appending with the first slice of the 20 slices; the $G^2$ of $\mathcal{G}1$ is given by $G^1$ appending with the second slice of the 20 slices; and so on. 

It is worth mentioning that there is no suitable well preprocessed dynamic networks to investigate the robustness of DNE to different DoCs. To achieve the goal, we utilize the unified preprocessing approach above to fairly deal with each real-world dataset, so as to generate several dynamic networks with different DoCs for each dataset.

\begin{table}[t]
  \centering
  \caption{The statistics of datasets and generated dynamic networks.}
  \renewcommand\tabcolsep{2.2pt}
  \scalebox{0.95}{
  \begin{threeparttable}
    \begin{tabular}{lccccc}
    \toprule
          & \multicolumn{1}{l}{DNC-Email} & \multicolumn{1}{l}{College-Msg} & \multicolumn{1}{l}{Co-Author} & \multicolumn{1}{l}{FB-Wall} & \multicolumn{1}{l}{Wiki-Talk} \\
    \midrule
    $\mathcal{G}$1-20slices & 170.4 & 531.7 & 4269.9 & 1213.6 & 3951.4 \\
    $\mathcal{G}$2-40slices & 85.2  & 265.8 & 2135.0 & 606.8 & 1975.7 \\
    $\mathcal{G}$3-60slices & 56.8  & 177.2 & 1423.3 & 404.5 & 1317.1 \\
    $\mathcal{G}$4-80slices & 42.6  & 132.9 & 1067.5 & 303.4 & 987.9 \\
    $\mathcal{G}$5-100slices & 34.1  & 106.3 & 854.0 & 242.7 & 790.3 \\
    \midrule
    $|\mathcal{E}|^{init}$ & 959   & 3202  & 21298 & 7420  & 19602 \\
    $|\mathcal{E}|^{last}$ & 4366  & 13835 & 106696 & 31691 & 98630 \\
    $|\mathcal{V}|^{init}$ & 555   & 755   & 1842  & 4343  & 6782 \\
    $|\mathcal{V}|^{last}$ & 1833  & 1893  & 4466  & 10300 & 31263 \\
    \bottomrule
    \end{tabular}%
    \begin{tablenotes} 
    \footnotesize 
		\item[1] The upper block presents Degree of Changes (DoCs), as defined in Definition 4, for each generated dynamic network.
		\item[2] The lower block shows the number of edges and nodes in the initial and last snapshots, which should be the same for the five dynamic networks generated from the same dataset.
    \end{tablenotes}
    \end{threeparttable}
    }
  \label{Tab1}%
\end{table}%

Apart from real-world datasets, we also prepare \textit{synthetic} datasets via the Barabási–Albert (BA) model \cite{barabasi1999emergence} to verify the scalability of the proposed method. The BA model is used to generate the dynamic network such that the node degree (w.r.t. counts) distribution obeys the power law distribution (i.e., a scale-free network) at each timestep. Specifically, it implements the preferential attachment mechanism to successively add a new node with $m_\text{BA}$ edges such that each edge connects the new node with higher probability to an existing node with larger node degree. The size of networks $|\mathcal{V}_\text{BA}|$ grows from a few nodes to \textit{millions} of nodes. The details of scalability test are presented and discussed in Section \ref{Sec5.4}

\subsection{Compared Methods}
The proposed method SG-EDNE (or EDNE for short) is compared to six state-of-the-art DNE methods. According to the literature review in Section \ref{literature}, BCGD-G and BGCG-L \cite{zhu2016scalable} belong to the matrix factorization based DNE; DynLINE \cite{du2018dynamic}, GloDyNE \cite{hou2020glodyne} and SG-EDNE (this work) belong to the Skip-Gram based DNE; DynTriad \cite{zhou2018dynamic} and tNEmbed \cite{singer2019node} employ other approaches to handle the DNE problem.

The original source codes of BCGD\footnote{\url{https://github.com/linhongseba/Temporal-Network-Embedding}},  DynLINE\footnote{ \url{https://github.com/lundu28/DynamicNetworkEmbedding}}, DynTriad\footnote{\url{https://github.com/luckiezhou/DynamicTriad}},  tNEmbed\footnote{\url{https://github.com/urielsinger/tNodeEmbed}}, and GloDyNE\footnote{\url{https://github.com/houchengbin/GloDyNE}} are used in the experiments. In tNEmbed, the link prediction architecture is used to obtain node embeddings, since the datasets in this work do not have node labels. In GloDyNE, the parameter retains 0.1 by default to balance effectiveness and efficiency. BCGD-G and BCGD-L are two proposed algorithms in BCGD for the type 2 and 4 algorithms respectively. Regarding other parameters, we search \{1e-4,1e-3,1e-2,1e-1\} for 'l' in BCGD-G and BCGD-L; \{3e-4,3e-3,3e-2,3e-1\} for 'learn-rate' in DynLINE; \{1e-3,1e-2,1e-1,1\} and \{1e-3,1e-2,1e-1,1\} for 'beta-triad' and 'beta-smooth' in DynTraid; and \{1,10,100,1000\} for 'train-skip' in tNEmbed. We find the recommended parameters in the original source codes obtain better results in most cases (among the 30 cases: DNC-Email and College-Msg datasets, 5 dynamic networks per dataset, and 3 downstream tasks), and are thus adopted for all experiments.

Without otherwise specified, for all experiments, SG-EDNE employs following default parameters: number of base learners $M=5$ and maximum restart probability $R^\text{max}=0.1$. For the parameters of Skip-Gram approach, we follow \cite{perozzi2014deepwalk,grover2016node2vec,hou2020glodyne} so that walks per node $r$, walk length $l$, window size $s$, and negative samples $q$ are set to 10, 80, 10, and 5 respectively.

For the fair comparison, the dimensionality of node embeddings for all methods is set to 128. Besides, all experiments are conducted given the following specification: Linux 4.15.0, 16 Intel-Xeon CPUs @2.20GHz, 256G memory, and Tesla P100 GPU with 16G memory. 


\subsection{Evaluation}
\begin{figure}[htbp]
    \centering
    \setlength{\abovecaptionskip}{-5pt}
    \includegraphics[width=0.489\textwidth]{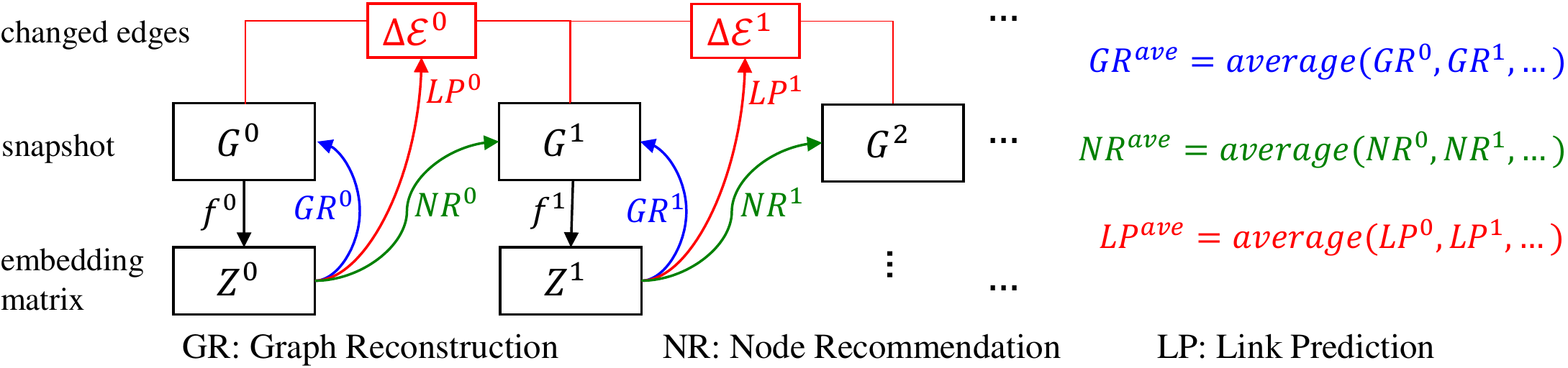}
    \caption{Evaluation protocol. For a task (e.g., $GR$), a method (e.g., SG-EDNE), and a dynamic network of a dataset (e.g., $\mathcal{G}1$ of DNC-Email), one result at each timestep $t$ (e.g., $GR^t$) is measured by an index (e.g., $MAP@5$). For the dynamic networks $\mathcal{G}1$, $\mathcal{G}2$, $\mathcal{G}3$, $\mathcal{G}4$, and $\mathcal{G}5$ generated from the same dataset, there are about 20, 40, 60, 80, and 100 timesteps respectively. Each point in Figure \ref{Fig4}-\ref{Fig6} depicts the averaged result over all timesteps (e.g., $GR^{avg}$), and we report the mean (e.g., of $GR^{avg}$) over 10 independent runs.}
    \label{Fig3}
\end{figure}

The node embeddings at each timestep are taken out and stored in the same format for all methods, so that we can utilize the same evaluation protocol as shown in Figure \ref{Fig3}. 

Three types of downstream tasks are employed to evaluate the quality of node embeddings. Graph Reconstruction (GR) tasks \cite{hou2020glodyne,zhu2018high,ma2020community} use current embeddings to reconstruct snapshot (by retrieving node neighbors for every node) at \textit{current} timestep. Node Recommendation (NR) tasks use current embeddings to recommend node neighbors for those affected nodes (by verifying their old and new neighbors) at \textit{next} timestep. Link Prediction (LP) tasks \cite{zhu2016scalable,zhou2018dynamic,hou2020glodyne} use current embeddings to predict new links (i.e., changed edges) at \textit{next} timestep. For GR and NR tasks, we adopt cosine similarity between embeddings for ranking, and employ mean average precision at k (MAP@k) \cite{goyal2017dyngem,wang2016structural} as the index. For LP tasks, we train an incremental logistic regression using previous new edges as positive samples and randomly sampled equal non-edges as negative samples, and employ area under the ROC curve (AUC) \cite{zhu2016scalable,yu2020node} as the index. The \textit{testing set} includes positive samples from future new edges at next timestep and equal negative samples from non-edges at next timestep. The edge features are given by Weighed-L1 and Weighed-L2 binary operators between node embeddings \cite{grover2016node2vec}. 

\section{Experiments and Results} \label{experiments}
\subsection{Comparative Study} \label{Sec5.1}
The comparative study intends to compare the effectiveness and robustness of the DNE methods, which include the six state-of-the-art DNE methods and proposed method SG-EDNE (or EDNE for short). In this work, the robustness is w.r.t. DoCs and is measured by the standard deviation (or stdev) over the five effectiveness results for the five input dynamic networks with different DoCs to a DNE method. 

\begin{figure}[!t]
    \centering
    \setlength{\abovecaptionskip}{-5pt}
    \includegraphics[width=0.489\textwidth]{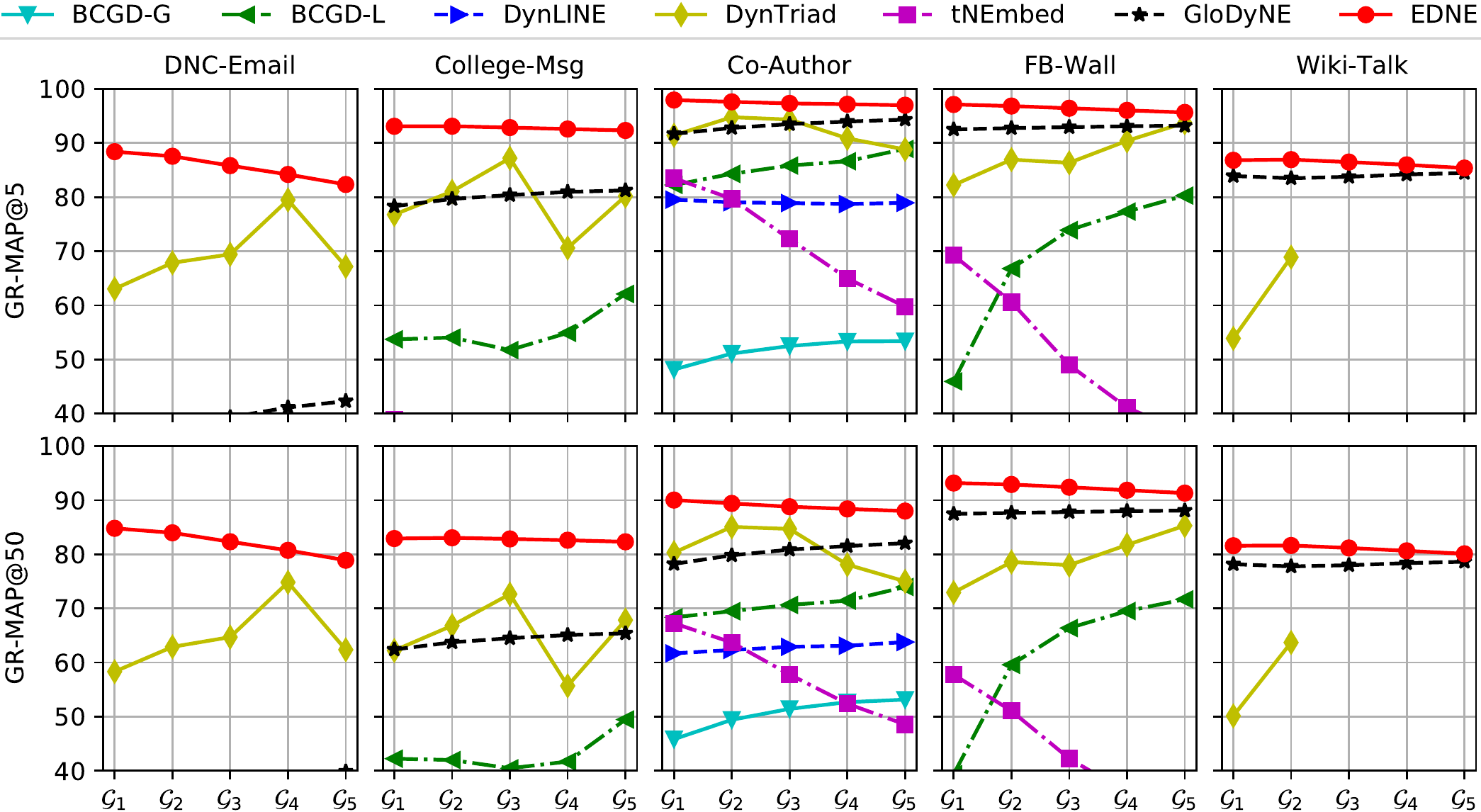}
    \caption{Comparative study for GR tasks. The evaluation protocol is illustrated in Figure \ref{Fig3}. Each point depicts the mean over 10 independent runs of the averaged result over all timesteps, i.e., $GR^{avg}$. For each dataset, the ranking of DoCs of five input dynamic networks is $\mathcal{G}1>\mathcal{G}2>\mathcal{G}3>\mathcal{G}4>\mathcal{G}5$.}
    \label{Fig4}
\end{figure}
\begin{figure}[!t]
    \centering
    \setlength{\abovecaptionskip}{-5pt}
    \includegraphics[width=0.489\textwidth]{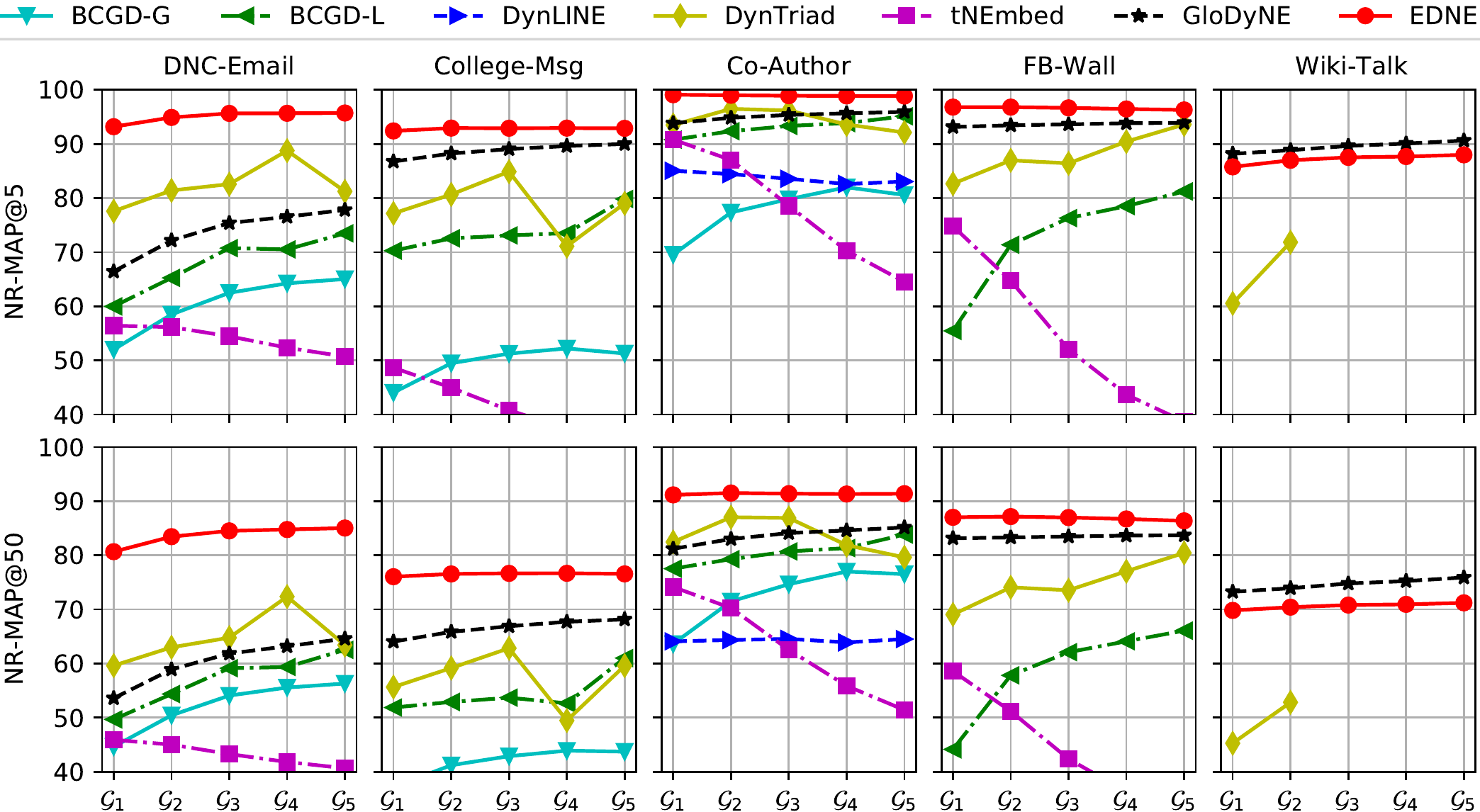}
    \caption{Comparative study for NR tasks.}
    \label{Fig5}
\end{figure}
\begin{figure}[!t]
    \centering
    \setlength{\abovecaptionskip}{-5pt}
    \includegraphics[width=0.489\textwidth]{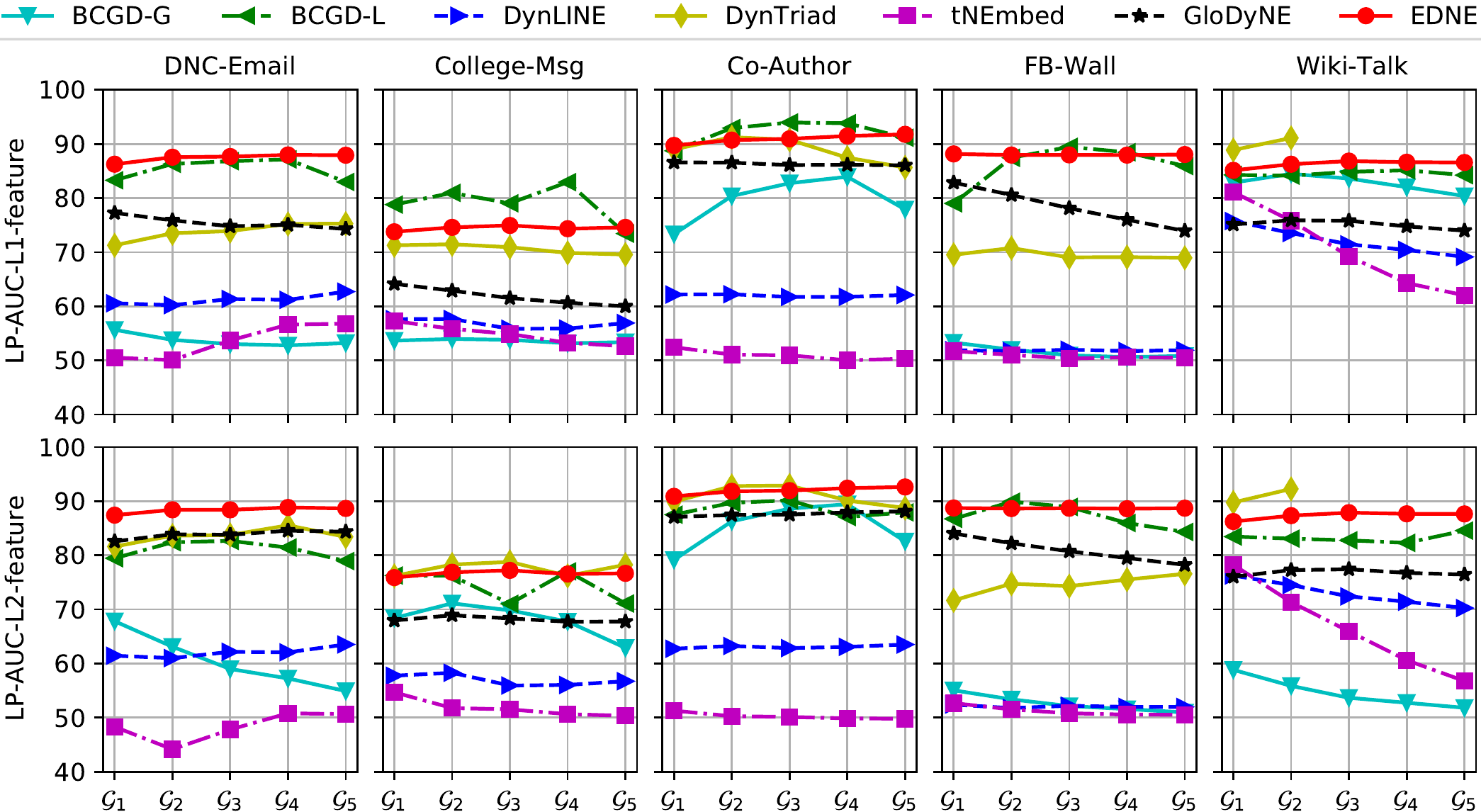}
    \caption{Comparative study for LP tasks.}
    \label{Fig6}
\end{figure}

First, we observe the quantitative results from Table \ref{Tab2} that the proposed method SG-EDNE achieves the best effectiveness and robustness (see mean$\pm$stdev) for most cases (25/30). The exceptions are GloDyNE in two NR tasks on Wiki-Talk, BCGD-L in two LP tasks on College-Msg, and BCGD-L in LP-AUC-L1-feature on Co-Author. Nevertheless, SG-EDNE is more robust w.r.t. DoCs (see stdev) in these exceptions, and still achieves the second best effectiveness (see mean). The same findings can be visually observed in Figure \ref{Fig4}-\ref{Fig6}.

Second, according to Figure \ref{Fig4}-\ref{Fig6} (effectivness below 40\% are omitted), we observe that DoCs would greatly affect the effectiveness of some DNE methods such as DynTriad in GR tasks on College-Msg, tNEmbed in NR tasks on Co-Author, and BCGD-L in LP tasks on FB-Wall. 

Third, it is interesting to observe there is no clear tendency that DNE methods prefer an input dynamic network with the smaller DoCs, i.e., the smoother changes, which is however assumed in several existing DNE works \cite{zhu2016scalable,li2017attributed,zhu2018high,zhang2018timers,ma2020community,du2018dynamic,zhou2018dynamic,goyal2017dyngem}.

Moreover, one more observation is that the performance of DNE methods might vary a lot for different datasets and different types of tasks. The reason is that different datasets and different types of tasks have different properties. For different datasets, the global (or average) clustering coefficient \cite{kemper2009valuation} of the last snapshot of DNC-Email is about 0.215, while that of Co-Author is about 0.665, and that of Wiki-Talk is about 0.114. For different types of tasks, GR using current embeddings to reconstruct current snapshot is a relative static task compared to LP for predicting future edges. Nevertheless, for the various datasets and tasks, the proposed method with default parameters can effectively and robustly outperforms other DNE methods for most cases.

\begin{table}[htbp]
  \centering
  \caption{Quantitative results of comparative study for all tasks.}
  \renewcommand\tabcolsep{3pt}
  \scalebox{0.85}{
  \begin{threeparttable}
    \begin{tabular}{lrrrrr}
    \toprule
          & \multicolumn{1}{c}{DNC-Email} & \multicolumn{1}{c}{College-Msg} & \multicolumn{1}{c}{Co-Author} & \multicolumn{1}{c}{FB-Wall} & \multicolumn{1}{c}{Wiki-Talk} \\
    \midrule
    \multicolumn{6}{c}{GR-MAP@5} \\
    \midrule
    BCGD-G & 13.77$\pm$0.44 & 26.42$\pm$0.79 & 51.68$\pm$1.95 & 0.73$\pm$0.02 & 0.53$\pm$0.01 \\
    BCGD-L & 22.57$\pm$0.78 & 55.3$\pm$3.55 & 85.57$\pm$2.24 & 68.84$\pm$12.3 & 3.37$\pm$1.09 \\
    DynLINE & 2.64$\pm$0.11 & 2.79$\pm$0.07 & 79.02$\pm$0.28 & 5.42$\pm$0.10 & 0.22$\pm$0.02 \\
    DynTriad & \textbf{69.39$\pm$5.46} & 79.15$\pm$5.44 & 92.03$\pm$2.24 & 87.92$\pm$3.90 & OOM \\
    tNEmbed   & 16.15$\pm$1.18 & 31.85$\pm$4.59 & 72.03$\pm$8.83 & 51.3$\pm$12.11 & 5.36$\pm$1.55 \\
    GloDyNE & 38.19$\pm$3.78 & \textbf{80.11$\pm$1.05} & \textbf{93.24$\pm$0.94} & \textbf{92.89$\pm$0.24} & \textbf{83.95$\pm$0.34} \\
    SG-EDNE & $^{\dagger}$\textbf{85.65$\pm$2.20} & $^{\dagger}$\textbf{92.78$\pm$0.30} & $^{\dagger}$\textbf{97.39$\pm$0.34} & $^{\dagger}$\textbf{96.40$\pm$0.53} & $^{\dagger}$\textbf{86.30$\pm$0.58} \\
    \midrule
    \multicolumn{6}{c}{GR-MAP@50} \\
    \midrule
    BCGD-G & 17.49$\pm$0.67 & 27.05$\pm$1.12 & 50.51$\pm$2.67 & 1.43$\pm$0.03 & 0.52$\pm$0.00 \\
    BCGD-L & 21.29$\pm$0.78 & 43.18$\pm$3.20 & 70.78$\pm$1.90 & 61.31$\pm$11.7 & 2.90$\pm$0.95 \\
    DynLINE & 2.42$\pm$0.06 & 2.97$\pm$0.06 & 62.76$\pm$0.71 & 5.45$\pm$0.06 & 0.25$\pm$0.02 \\
    DynTriad & \textbf{64.62$\pm$5.51} & \textbf{65.05$\pm$5.73} & \textbf{80.63$\pm$3.86} & 79.31$\pm$4.11 & OOM \\
    tNEmbed   & 14.30$\pm$1.08 & 25.95$\pm$3.17 & 57.94$\pm$6.91 & 43.91$\pm$9.39 & 4.66$\pm$1.18 \\
    GloDyNE & 36.15$\pm$3.41 & 64.23$\pm$1.07 & 80.49$\pm$1.36 & \textbf{87.80$\pm$0.23} & \textbf{78.18$\pm$0.30} \\
    SG-EDNE & $^{\dagger}$\textbf{82.14$\pm$2.14} & $^{\dagger}$\textbf{82.74$\pm$0.26} & $^{\dagger}$\textbf{88.92$\pm$0.72} & $^{\dagger}$\textbf{92.32$\pm$0.68} & $^{\dagger}$\textbf{81.02$\pm$0.59} \\
    \midrule
    \multicolumn{6}{c}{NR-MAP@5} \\
    \midrule
    BCGD-G & 60.45$\pm$4.78 & 49.65$\pm$2.94 & 77.85$\pm$4.41 & 1.52$\pm$0.10 & 7.02$\pm$1.28 \\
    BCGD-L & 67.99$\pm$4.81 & 73.85$\pm$3.19 & 93.08$\pm$1.45 & 72.57$\pm$9.15 & 24.73$\pm$5.90 \\
    DynLINE & 14.80$\pm$0.74 & 5.05$\pm$0.29 & 83.73$\pm$0.89 & 7.99$\pm$0.14 & 1.37$\pm$0.09 \\
    DynTriad & \textbf{82.31$\pm$3.64} & 78.56$\pm$4.52 & 94.37$\pm$1.68 & 88.01$\pm$3.72 & OOM \\
    tNEmbed   & 54.02$\pm$2.20 & 41.45$\pm$4.84 & 78.24$\pm$9.87 & 54.8$\pm$13.34 & 23.90$\pm$5.49 \\
    GloDyNE & 73.67$\pm$4.08 & \textbf{88.72$\pm$1.15} & \textbf{95.12$\pm$0.75} & \textbf{93.58$\pm$0.28} & $^{\dagger}$\textbf{89.47$\pm$0.87} \\
    SG-EDNE & $^{\dagger}$\textbf{95.02$\pm$0.97} & $^{\dagger}$\textbf{92.80$\pm$0.20} & $^{\dagger}$\textbf{98.94$\pm$0.09} & $^{\dagger}$\textbf{96.60$\pm$0.20} & \textbf{87.18$\pm$0.78} \\
    \midrule
    \multicolumn{6}{c}{NR-MAP@50} \\
    \midrule
    BCGD-G & 52.20$\pm$4.29 & 41.78$\pm$2.50 & 72.65$\pm$4.91 & 2.35$\pm$0.12 & 6.68$\pm$1.20 \\
    BCGD-L & 57.01$\pm$4.50 & 54.42$\pm$3.34 & 80.53$\pm$2.10 & 58.85$\pm$7.85 & 18.67$\pm$4.18 \\
    DynLINE & 11.75$\pm$0.30 & 5.32$\pm$0.17 & 64.28$\pm$0.25 & 7.95$\pm$0.05 & 1.67$\pm$0.11 \\
    DynTriad & \textbf{64.63$\pm$4.22} & 57.35$\pm$4.55 & 83.57$\pm$2.92 & 74.83$\pm$3.79 & OOM \\
    tNEmbed   & 43.33$\pm$1.94 & 31.02$\pm$3.00 & 62.85$\pm$8.51 & 44.22$\pm$9.54 & 18.62$\pm$3.62 \\
    GloDyNE & 60.42$\pm$3.90 & \textbf{66.53$\pm$1.47} & \textbf{83.61$\pm$1.40} & \textbf{83.45$\pm$0.21} & $^{\dagger}$\textbf{74.62$\pm$0.94} \\
    SG-EDNE & $^{\dagger}$\textbf{83.68$\pm$1.60} & $^{\dagger}$\textbf{76.50$\pm$0.23} & $^{\dagger}$\textbf{91.35$\pm$0.11} & $^{\dagger}$\textbf{86.84$\pm$0.28} & \textbf{70.63$\pm$0.49} \\
    \midrule
    \multicolumn{6}{c}{LP-AUC-L1-feature} \\
    \midrule
    BCGD-G & 53.69$\pm$1.04 & 53.59$\pm$0.29 & 79.65$\pm$3.78 & 51.54$\pm$1.00 & 82.66$\pm$1.40 \\
    BCGD-L & \textbf{85.30$\pm$1.80} & $^{\dagger}$\textbf{79.03$\pm$3.17} & $^{\dagger}$\textbf{92.12$\pm$1.97} & \textbf{86.05$\pm$3.71} & \textbf{84.52$\pm$0.39} \\
    DynLINE & 61.21$\pm$0.86 & 56.78$\pm$0.80 & 61.99$\pm$0.21 & 51.84$\pm$0.10 & 72.05$\pm$2.34 \\
    DynTriad & 73.83$\pm$1.46 & 70.61$\pm$0.76 & 88.83$\pm$2.08 & 69.47$\pm$0.67 & OOM \\
    tNEmbed   & 53.53$\pm$2.88 & 54.77$\pm$1.70 & 50.96$\pm$0.83 & 50.83$\pm$0.49 & 70.49$\pm$7.10 \\
    GloDyNE & 75.45$\pm$1.04 & 61.84$\pm$1.51 & 86.29$\pm$0.23 & 78.29$\pm$3.17 & 75.11$\pm$0.71 \\
    SG-EDNE & $^{\dagger}$\textbf{87.47$\pm$0.63} & \textbf{74.44$\pm$0.39} & \textbf{90.92$\pm$0.70} & $^{\dagger}$\textbf{88.01$\pm$0.08} & $^{\dagger}$\textbf{86.28$\pm$0.61} \\
    \midrule
    \multicolumn{6}{c}{LP-AUC-L2-feature} \\
    \midrule
    BCGD-G & 60.41$\pm$4.57 & 68.00$\pm$2.81 & 85.21$\pm$3.85 & 52.62$\pm$1.45 & 54.59$\pm$2.51 \\
    BCGD-L & 80.97$\pm$1.53 & 74.32$\pm$2.68 & 88.47$\pm$1.22 & \textbf{87.16$\pm$2.02} & \textbf{83.23$\pm$0.77} \\
    DynLINE & 62.04$\pm$0.85 & 56.95$\pm$0.93 & 63.08$\pm$0.28 & 52.07$\pm$0.22 & 72.97$\pm$2.15 \\
    DynTriad & 83.58$\pm$1.24 & $^{\dagger}$\textbf{77.53$\pm$1.11} & \textbf{90.88$\pm$1.66} & 74.56$\pm$1.64 & OOM \\
    tNEmbed  & 48.34$\pm$2.42 & 51.79$\pm$1.54 & 50.28$\pm$0.54 & 51.22$\pm$0.81 & 66.58$\pm$7.63 \\
    GloDyNE & \textbf{83.83$\pm$0.69} & 68.13$\pm$0.45 & 87.61$\pm$0.37 & 80.94$\pm$2.03 & 76.78$\pm$0.51 \\
    SG-EDNE & $^{\dagger}$\textbf{88.34$\pm$0.49} & \textbf{76.63$\pm$0.44} & $^{\dagger}$\textbf{91.93$\pm$0.61} & $^{\dagger}$\textbf{88.67$\pm$0.05} & $^{\dagger}$\textbf{87.34$\pm$0.58} \\
    \bottomrule
    \end{tabular}%
    \begin{tablenotes} 
    \footnotesize 
		\item Each entry quantifies mean$\pm$stdev of each line (with five effectiveness scores) in Figure \ref{Fig4}, \ref{Fig5}, and \ref{Fig6}. The top two are in bold, and $^{\dagger}$ highlights the best one.
    \end{tablenotes}
    \end{threeparttable}
    }
  \label{Tab2}%
\end{table}%

\subsection{Ablation Study} \label{Sec5.2}
The ablation study aims to investigate if the key components in the proposed method (here termed as EDNE-rwr) are useful. EDNE-rw is a variant that replaces random walks with restart (rwr) used in EDNE-rwr with random walks (rw). EDNE-rw-fix is a variant that fixes one set of node sequences for all base models compared to EDNE-rw that generates different set of node sequences to each base model. DNE-rw is a variant that uses only one base model compared to EDNE-rw that employs more than one base models. EDNE-rwr-ws is a variant without the scaling operation compared to EDNE-rwr. The results for GR tasks are shown in Table \ref{Tab3}, and each entry is obtained in the same way as that in Table \ref{Tab2}. The results for NR and LP tasks can be found in Appendix.

\begin{table}[htbp]
  \centering
  \caption{Ablation study for GR tasks.}
  \renewcommand\tabcolsep{2.0pt}
  \scalebox{0.85}{
    \begin{tabular}{lrrrrr}
    \toprule
          & \multicolumn{1}{c}{DNC-Email} & \multicolumn{1}{c}{College-Msg} & \multicolumn{1}{c}{Co-Author} & \multicolumn{1}{c}{FB-Wall} & \multicolumn{1}{c}{Wiki-Talk} \\
    \midrule
    \multicolumn{6}{c}{GR-MAP@5} \\
    \midrule
    DNE-rw & 54.41$\pm$3.64 & 90.64$\pm$1.52 & 94.28$\pm$0.89 & 93.86$\pm$1.13 & $^{\dagger}$\textbf{90.47$\pm$2.64} \\
    EDNE-rw-fix & 59.48$\pm$2.65 & 85.97$\pm$0.96 & 94.49$\pm$0.73 & 93.12$\pm$0.87 & 43.10$\pm$0.87 \\
    EDNE-rw & \textbf{83.35$\pm$1.22} & \textbf{92.57$\pm$0.21} & \textbf{97.06$\pm$0.33} & \textbf{95.47$\pm$0.44} & 84.88$\pm$0.44 \\
    EDNE-rwr & $^{\dagger}$\textbf{85.65$\pm$2.20} & $^{\dagger}$\textbf{92.78$\pm$0.30} & $^{\dagger}$\textbf{97.39$\pm$0.34} & $^{\dagger}$\textbf{96.40$\pm$0.53} & \textbf{86.30$\pm$0.58} \\
    EDNE-rwr-ws & 82.29$\pm$3.32 & 92.17$\pm$0.78 & 96.96$\pm$0.49 & 94.89$\pm$0.75 & 85.49$\pm$0.93 \\
    \midrule
    \multicolumn{6}{c}{GR-MAP@50} \\
    \midrule
    DNE-rw & 51.04$\pm$3.53 & 80.68$\pm$2.27 & 84.91$\pm$1.63 & 90.53$\pm$1.84 & $^{\dagger}$\textbf{86.85$\pm$3.10} \\
    EDNE-rw-fix & 55.52$\pm$2.64 & 71.83$\pm$1.55 & 83.60$\pm$1.38 & 87.42$\pm$1.48 & 40.53$\pm$0.80 \\
    EDNE-rw & \textbf{79.75$\pm$1.22} & \textbf{82.36$\pm$0.20} & \textbf{88.34$\pm$0.66} & \textbf{91.44$\pm$0.55} & 79.58$\pm$0.46 \\
    EDNE-rwr & $^{\dagger}$\textbf{82.14$\pm$2.14} & $^{\dagger}$\textbf{82.74$\pm$0.26} & $^{\dagger}$\textbf{88.92$\pm$0.72} & $^{\dagger}$\textbf{92.32$\pm$0.68} & \textbf{81.02$\pm$0.59} \\
    EDNE-rwr-ws & 77.72$\pm$3.32 & 78.23$\pm$0.87 & 86.62$\pm$1.08 & 88.89$\pm$0.95 & 80.32$\pm$0.90 \\
    \bottomrule
    \end{tabular}%
    }
  \label{Tab3}%
\end{table}%

First, from the \textit{diversity} perspective of the training samples to ensembles, EDNE-rwr $>$ EDNE-rw $>$ EDNE-rw-fix. The results among them indicate that enhancing the diversity among base models can improve the performance. Second, comparing the results of EDNE-rwr to EDNE-rwr-ws, it demonstrates the usefulness of the scaling operation.

Third, regarding DNE-rw, it is interesting to observe that applying ensembles to DNE by EDNE-rw-fix strategy would lead to the worse results, while other carefully designed ensembles by EDNE-rw strategy and EDNE-rwr strategy often obtain the better results. One exception is that the proposed method by EDNE-rwr on Wiki-Talk is outperformed by DNE-rw without ensembles, though EDNE-rwr is more robust. This encourages us to further exploit a better design of ensembles for DNE methods as a future work.

\subsection{Parameter Sensitivity} \label{Sec5.3} 
The parameter sensitivity analysis tries to investigate two important hyper-parameters $M$ and $R^\text{max}$. We vary $M$ from 1 to 10 with step 1 and $R^\text{max}$ from 0.1 to 0.9 with step 0.2. There are totally $10\times 5$ points, and each point is given by the same evaluation protocol as illustrated in Figure \ref{Fig3}. The results for GR tasks are shown in Figure \ref{Fig7}, while the results for NR and LP tasks can be found in Appendix.

\begin{figure}[htbp]
    \centering
    \setlength{\abovecaptionskip}{-3pt}
    \includegraphics[width=0.489\textwidth]{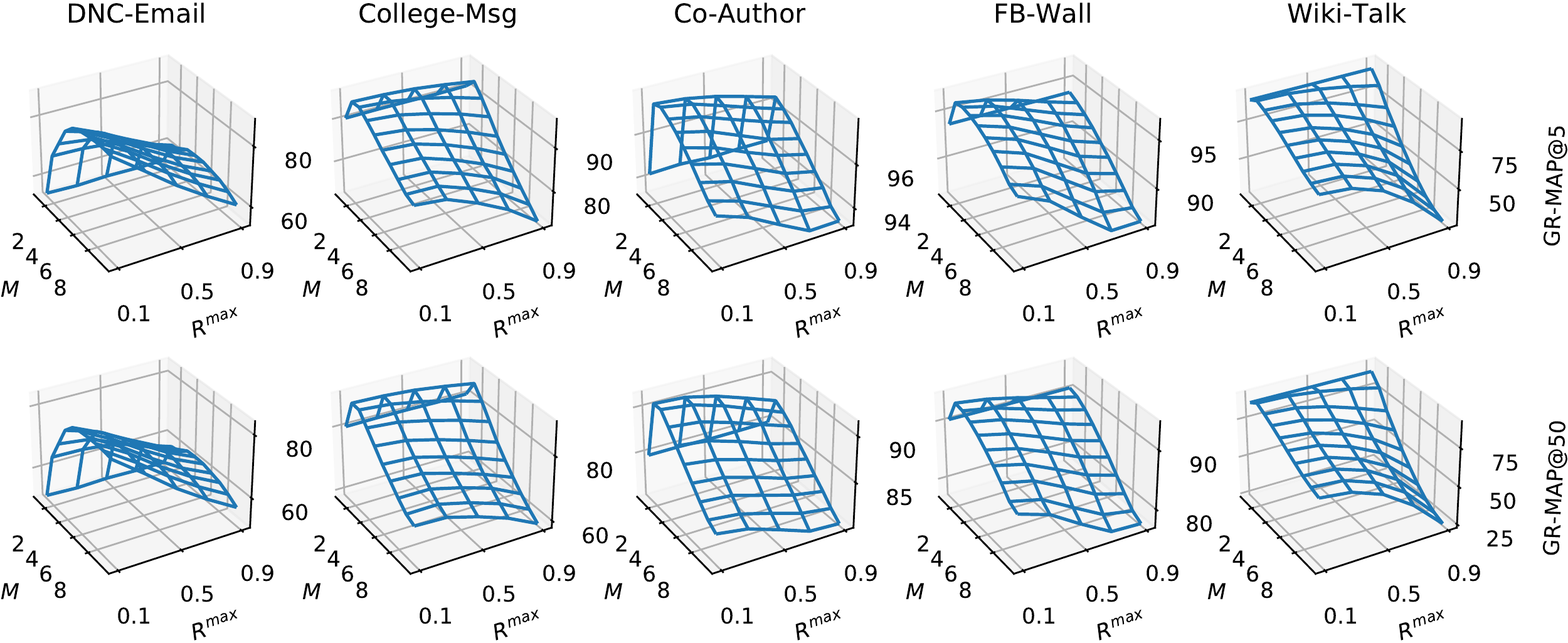}
    \caption{Parameter sensitivity for GR tasks. $M$ is for the number of base models and $R^\text{max}$ is for the maximum restart probability.}
    \label{Fig7}
\end{figure}

\begin{figure}[htbp]
    \centering
    \includegraphics[width=0.48\textwidth]{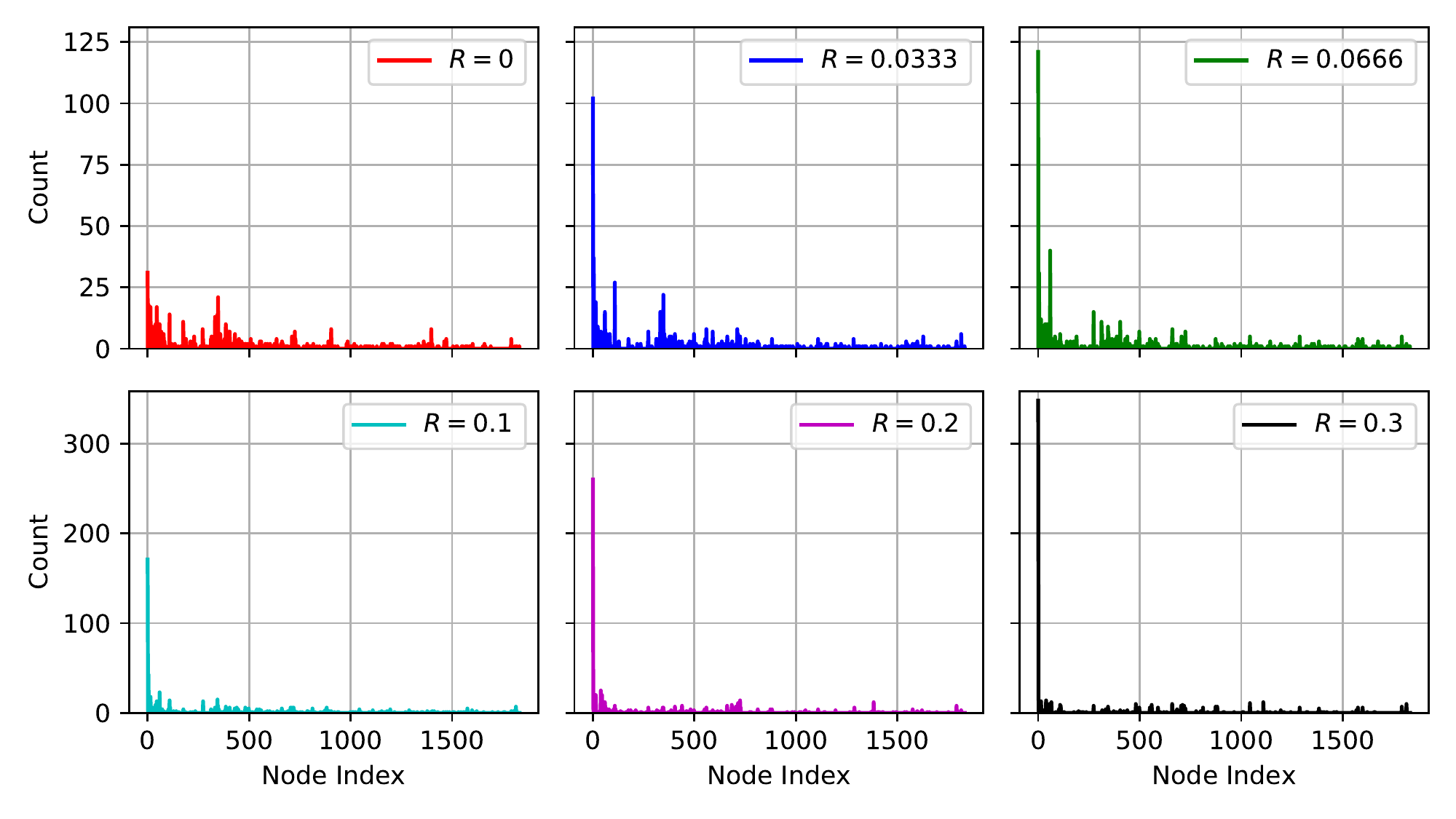}
    \caption{Statistics of RWR with different restart probabilities $R$ starting from the same node on the last snapshot of DNC-Email.}
    \label{Fig8}
\end{figure}

First, for the number of base models (or base learners) $M$, increasing $M$ would not always improve the performance, though $M>1$ often achieves better performance. It seems that different datasets have quite different preferences in $M$. Second, for the maximum restart probability $R^\text{max}$, a smaller $R^\text{max}$ is often preferred in most cases. Recall Eq. (\ref{eq9}) for assigning different restart probabilities to RWR based on $R^\text{max}$, which implies that the proposed method employs a smaller restart probability than $R^\text{max}$ in each base model. This motivates us to study the property of RWR with different restart probabilities $R$, which is shown in Figure \ref{Fig8}. 

For Figure \ref{Fig8}, we randomly pick a node and conduct RWR with different $R$ starting from the same node on the last snapshot of DNC-Email. We observe that different $R$s can explore different levels of local-global neighbors (i.e., different levels of local-global topology) of the starting node. In particular, a smaller $R$ explores a richer set of neighbors (more evenly visit various nodes), while as $R$ becomes larger, the \textit{diversity} of the the neighboring information being explored is reduced (more likely revisit few nodes). This might be the reason why the proposed method prefers a smaller $R^\text{max}$. 

\subsection{Scalability} \label{Sec5.4} 
This section presents the scalability test for SG-EDNE to verify its theoretical complexity given in Section \ref{complexity}. Since SG-EDNE consists of two stages namely the offline stage (at initial timestep or when retraining is needed) and the online stage (during incremental learning), two sets of datasets are generated via the BA model (see Section \ref{Sec4.1}) with $m_\text{BA}=4$ new edges per new node. For the offline stage, the number of nodes $|\mathcal{V}_\text{BA}|$ in each network ranges from $2^6$ to $2^{20}$ with step 1 in the exponent. For the online stage, the number of newly emerging nodes $|\Delta \mathcal{V}_\text{BA}|$ per snapshot ranges from $2^2$ to $2^{16}$ (lasting for 20 snapshots so the last snapshot of a dynamic network has about $|\Delta \mathcal{V}_\text{BA}| \times 20$ nodes).

\begin{figure}[htbp]
    \centering
    \setlength{\abovecaptionskip}{-5pt}
    \includegraphics[width=0.489\textwidth]{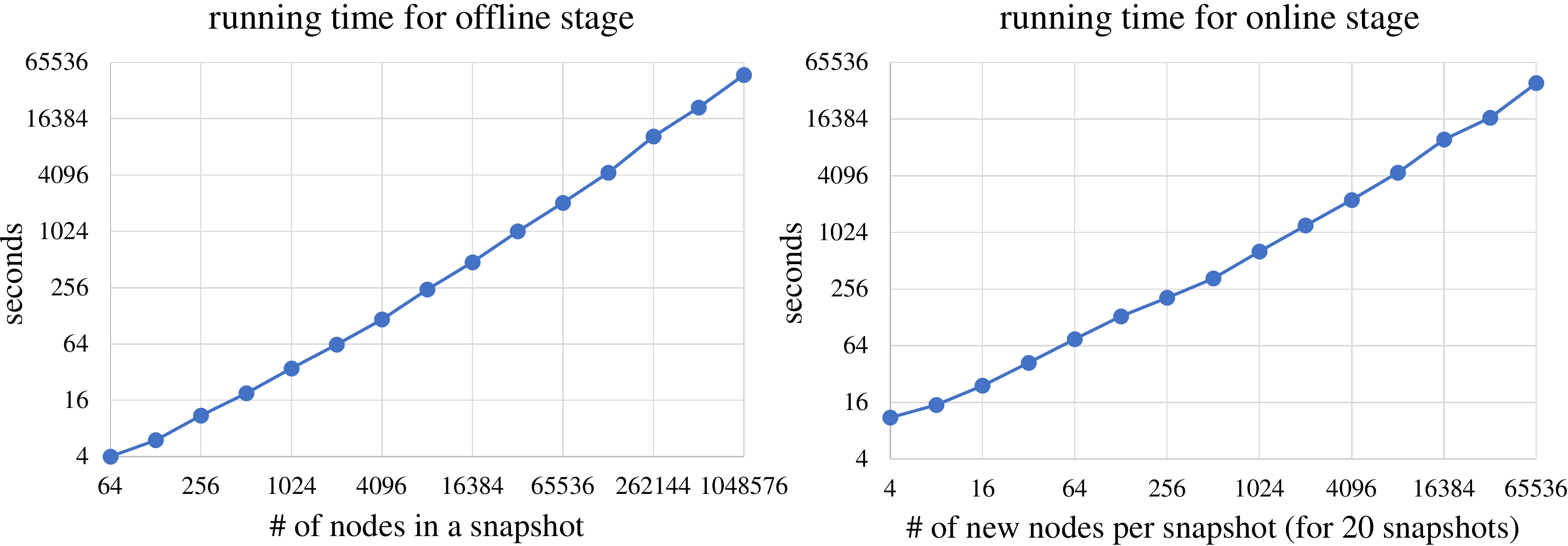}
    \caption{Scalability test for the proposed method. All axes are in $\log_2$ scale.}
    \label{Scalability}
\end{figure}

It might be easy to observe from Figure \ref{Scalability} that the results of scalability test are consistent with the theoretical complexity. The running time for offline stage and online stage is linearly proportional to $|\mathcal{V}|$ and $(|\mathcal{V}^t|+|\Delta\mathcal{V}^t|)$ respectively.

\section{Conclusion}
In this work, we suggested the robustness issue of DNE methods w.r.t. Degree of Changes (DoCs), and proposed an effective and more robust DNE method namely SG-EDNE basaed on the diversity enhanced ensembles of the incremental Skip-Gram model. The comparative study revealed that the existing DNE methods are not robust enough to DoCs, and also demonstrated the superior effectiveness and robustness of SG-EDNE compared to the existing DNE methods. The further empirical studies for SG-EDNE verified the benefits of special designs in SG-EDNE and its scalability. The source code is available at \url{https://github.com/houchengbin/SG-EDNE}

We acknowledge that our first attempt to the robustness issue of DNE w.r.t. DoCs remains limitations. One can also preprocess a dynamic network with nonuniform number of streaming edges per timestep, and then study the robustness of DNE under this scenario. It might be another interesting but probably a more challenging scenario for the future work. Moreover, it is also interesting to develop a better design of DNE ensembles, e.g., adopting other embedding model(s) as base learners, or exploiting a better strategy to enhance the diversity among base learners.



\section*{Acknowledgment}
This work was supported in part by the Natural Science Foundation of China under Grants 61672478 and 61806090, in part by the Guangdong Provincial Key Laboratory under Grant 2020B121201001, in part by the Program for Guangdong Introducing Innovative and Entrepreneurial Teams under Grant 2017ZT07X386, in part by the Shenzhen Peacock Plan under Grant KQTD2016112514355531, and in part by the Stable Support Plan Program of Shenzhen Natural Science Fund Grant 20200925154942002. The authors would also like to thank the support from the Huawei-SUSTech AI RAMS Joint Innovation Laboratory.

\appendices
\section{Results of Ablation Study (CONT'D)} \label{C.1}
In addition to Table \ref{Tab3} (for GR tasks), we provide the results of ablation study for NR and LP tasks as shown in Table \ref{Tab4}. 

\begin{table}[htbp]
  \centering
  \caption{Ablation study for NR and LP tasks.}
  \renewcommand\tabcolsep{2.0pt}
  \scalebox{0.82}{
    \begin{tabular}{lrrrrr}
    \toprule
          & \multicolumn{1}{c}{DNC-Email} & \multicolumn{1}{c}{College-Msg} & \multicolumn{1}{c}{Co-Author} & \multicolumn{1}{c}{FB-Wall} & \multicolumn{1}{c}{Wiki-Talk} \\
    \midrule
    \multicolumn{6}{c}{NR-MAP@5} \\
    \midrule
    DNE-rw & 81.22$\pm$0.71 & 88.98$\pm$1.85 & 96.00$\pm$0.99 & 94.41$\pm$1.02 & \textbf{92.62$\pm$1.52} \\
    EDNE-rw-fix & 87.90$\pm$1.10 & 87.46$\pm$0.59 & 97.18$\pm$0.43 & 93.56$\pm$0.67 & 70.97$\pm$1.11 \\
    EDNE-rw & \textbf{94.24$\pm$1.33} & 92.34$\pm$0.21 & 98.67$\pm$0.15 & \textbf{95.82$\pm$0.24} & 85.53$\pm$0.71 \\
    EDNE-rwr & \textbf{95.02$\pm$0.97} & \textbf{92.80$\pm$0.20} & \textbf{98.94$\pm$0.09} & \textbf{96.60$\pm$0.20} & 87.18$\pm$0.78 \\
    EDNE-rwr-ws & 93.62$\pm$0.85 & \textbf{94.44$\pm$0.19} & \textbf{98.69$\pm$0.18} & 95.09$\pm$0.38 & \textbf{87.64$\pm$0.79} \\
    \midrule
    \multicolumn{6}{c}{NR-MAP@50} \\
    \midrule
    DNE-rw & 67.46$\pm$0.59 & 71.15$\pm$2.88 & 87.43$\pm$1.30 & 86.72$\pm$1.87 & \textbf{81.14$\pm$3.35} \\
    EDNE-rw-fix & 74.62$\pm$1.22 & 67.12$\pm$1.20 & 87.70$\pm$0.60 & 82.10$\pm$1.22 & 54.54$\pm$0.71 \\
    EDNE-rw & \textbf{82.58$\pm$1.85} & \textbf{75.60$\pm$0.19} & \textbf{90.81$\pm$0.13} & \textbf{86.13$\pm$0.29} & 68.61$\pm$0.29 \\
    EDNE-rwr & \textbf{83.68$\pm$1.60} & \textbf{76.50$\pm$0.23} & \textbf{91.35$\pm$0.11} & \textbf{86.84$\pm$0.28} & 70.63$\pm$0.49 \\
    EDNE-rwr-ws & 80.53$\pm$1.28 & 74.93$\pm$0.29 & 89.32$\pm$0.30 & 82.89$\pm$0.52 & \textbf{71.33$\pm$0.66} \\
    \midrule
    \multicolumn{6}{c}{LP-AUC-L1-feature} \\
    \midrule
    DNE-rw & 84.86$\pm$0.60 & 72.49$\pm$0.34 & 87.91$\pm$0.33 & 84.35$\pm$0.49 & 77.14$\pm$0.94 \\
    EDNE-rw-fix & 86.21$\pm$0.44 & 73.35$\pm$0.41 & 90.43$\pm$0.54 & 87.29$\pm$0.11 & 85.16$\pm$0.58 \\
    EDNE-rw & \textbf{87.48$\pm$0.64} & \textbf{74.79$\pm$0.50} & \textbf{90.90$\pm$0.68} & \textbf{87.83$\pm$0.02} & \textbf{86.29$\pm$0.68} \\
    EDNE-rwr & \textbf{87.47$\pm$0.63} & \textbf{74.44$\pm$0.39} & \textbf{90.92$\pm$0.70} & \textbf{88.01$\pm$0.08} & \textbf{86.28$\pm$0.61} \\
    EDNE-rwr-ws & 85.16$\pm$1.11 & 70.28$\pm$0.25 & 89.36$\pm$0.23 & 84.61$\pm$1.37 & 84.96$\pm$0.49 \\
    \midrule
    \multicolumn{6}{c}{LP-AUC-L2-feature} \\
    \midrule
    DNE-rw & 85.96$\pm$0.48 & 75.01$\pm$0.40 & 89.17$\pm$0.41 & 85.69$\pm$0.27 & 78.79$\pm$0.66 \\
    EDNE-rw-fix & 87.13$\pm$0.32 & 75.48$\pm$0.47 & 91.34$\pm$0.49 & 87.92$\pm$0.07 & 86.13$\pm$0.57 \\
    EDNE-rw & \textbf{88.32$\pm$0.53} & \textbf{77.01$\pm$0.58} & \textbf{91.89$\pm$0.58} & \textbf{88.45$\pm$0.03} & \textbf{87.37$\pm$0.66} \\
    EDNE-rwr & \textbf{88.34$\pm$0.49} & \textbf{76.63$\pm$0.44} & \textbf{91.93$\pm$0.61} & \textbf{88.67$\pm$0.05} & \textbf{87.34$\pm$0.58} \\
    EDNE-rwr-ws & 85.14$\pm$1.00 & 70.29$\pm$0.27 & 89.54$\pm$0.20 & 83.24$\pm$2.17 & 85.31$\pm$0.51 \\
    \bottomrule
    \end{tabular}%
    }
  \label{Tab4}%
\end{table}%

The findings for NR and LP tasks are similar to that for GR tasks as discussed in Section \ref{Sec5.2}. There are two exceptions. First, EDNE-rwr-ws outperforms EDNE-rwr on College-Msg in NR-MAP@5 and on Wiki-Talk in both NR tasks. This finding indicates that the scaling operation, i.e., [0, 1] min-max scaling, may fail sometimes. However, for all other 27/30 cases, the scaling operation can improve the performance. 

Second, DNE-rw (without ensembles) obtains the worst results than all other variants of DNE ensembles in LP tasks, while DNE-rw obtains the best results in NR tasks GR tasks (see Table \ref{Tab3}). This observation may encourage us to employ the ensembles in LP tasks.

\section{Results of Parameter Sensitivity (CONT'D)}
\begin{figure}[htbp]
    \centering
    \setlength{\abovecaptionskip}{-3pt}
    \includegraphics[width=0.489\textwidth]{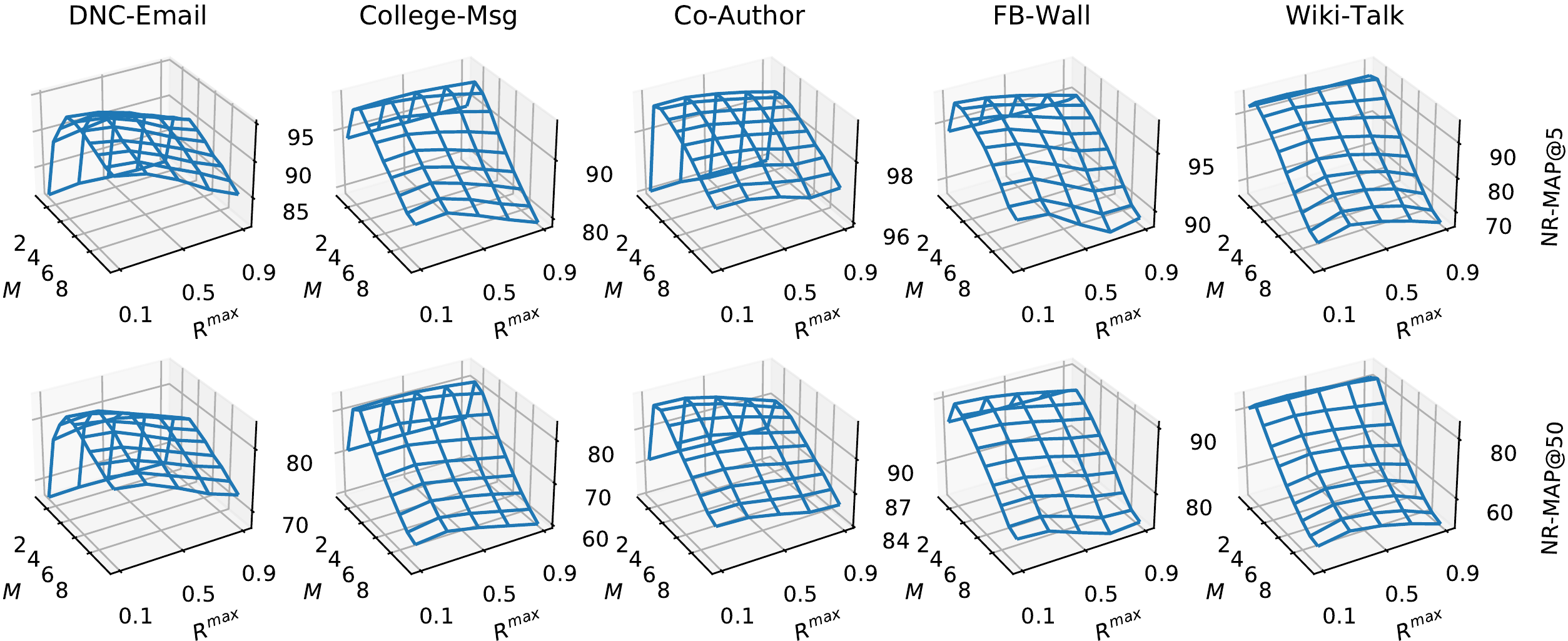}
    \caption{Parameter sensitivity for NR tasks. $M$ is for the number of base models and $R^\text{max}$ is for the maximum restart probability.}
    \label{Fig9}
\end{figure}

\begin{figure}[htbp]
    \centering
    \setlength{\abovecaptionskip}{-3pt}
    \includegraphics[width=0.489\textwidth]{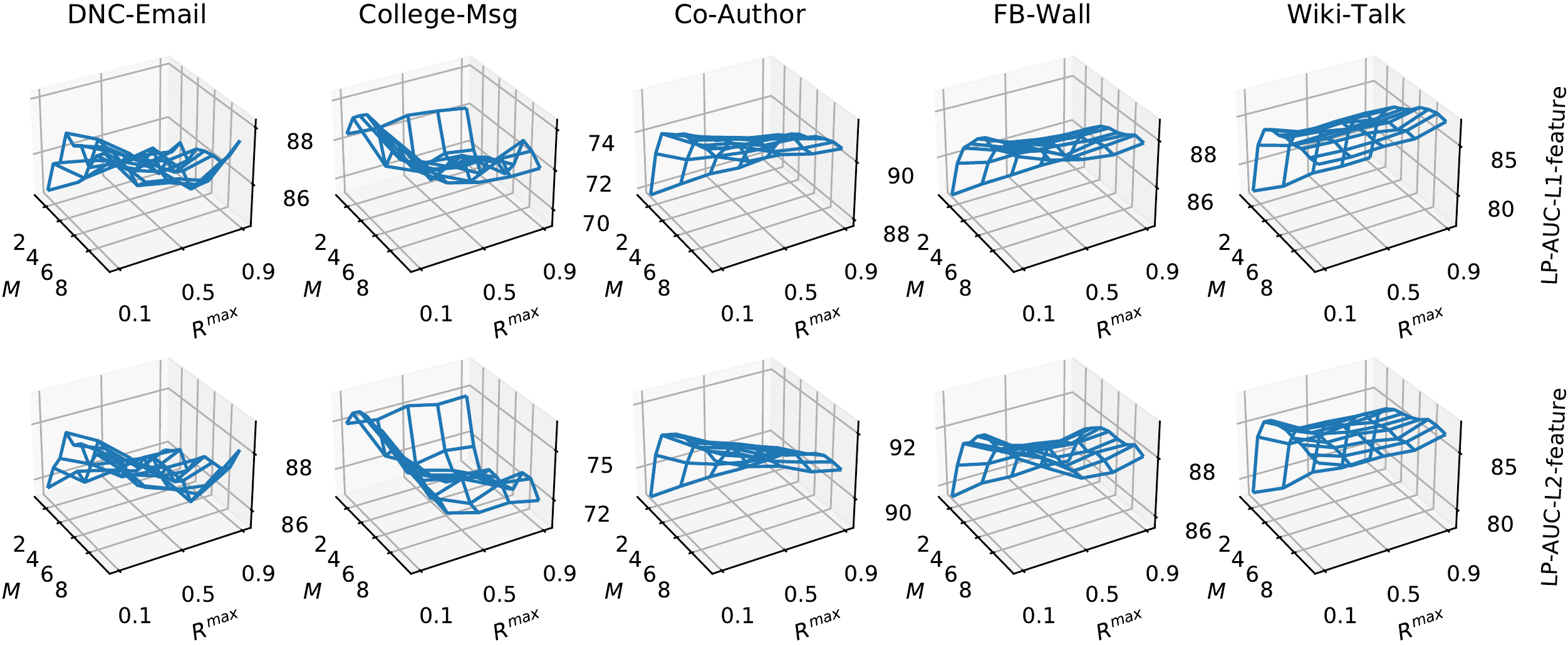}
    \caption{Parameter sensitivity for LP tasks. $M$ is for the number of base models and $R^\text{max}$ is for the maximum restart probability.}
    \label{Fig10}
\end{figure}

In addition to Figure \ref{Fig7} (for GR tasks), we provide the results of parameter sensitivity for NR and LP tasks as shown in Figure \ref{Fig9} and \ref{Fig10}. The findings for NR and LP tasks are the same to that for GR tasks as discussed in Section \ref{Sec5.3}. 

One new observation, according to Figure \ref{Fig10}, is that LP tasks often prefer a lager $M$ than GR and NR tasks. This observation again encourages us to employ the ensembles in LP tasks as also suggested in Appendix \ref{C.1} above.



\ifCLASSOPTIONcaptionsoff
  \newpage
\fi



\bibliographystyle{IEEEtran}
\bibliography{IEEEtran.bib}
%



%

\begin{IEEEbiography}[{\includegraphics[width=1in,height=1.25in,clip,keepaspectratio]{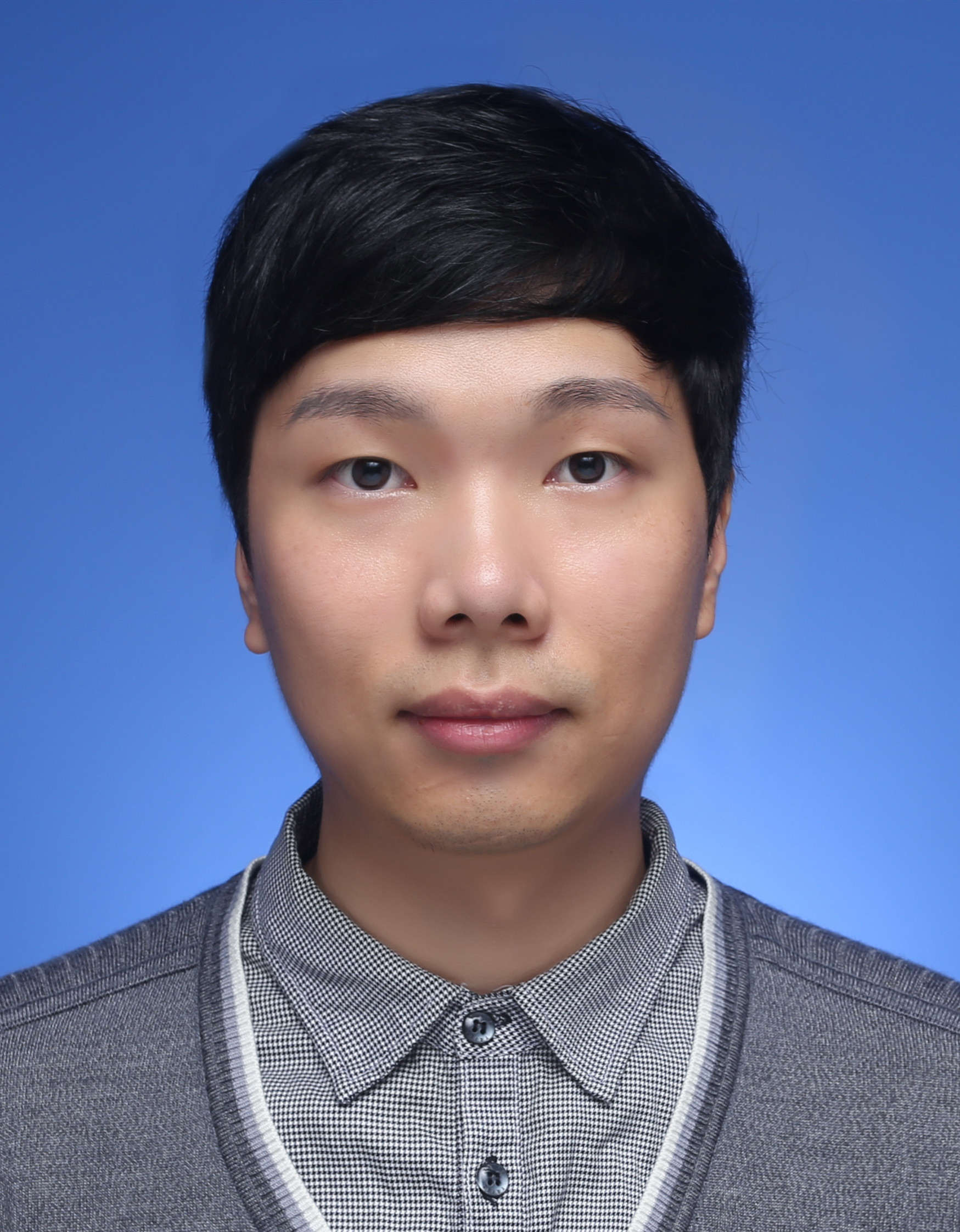}}]{Chengbin Hou} received the B.Eng. first class degree and the M.Sc. distinction degree from University of Liverpool (July 2014) and Imperial College London (November 2015) respectively. He started his Ph.D. in September 2017. Before his PhD study, he worked as an engineer at Huawei. Currently, he is pursuing his Ph.D. degree at Southern University of Science and Technology (SUSTech) and University of Birmingham. His research interests include machine learning and data mining on graph data.
\end{IEEEbiography}

\begin{IEEEbiography}[{\includegraphics[width=1in,height=1.25in,clip,keepaspectratio]{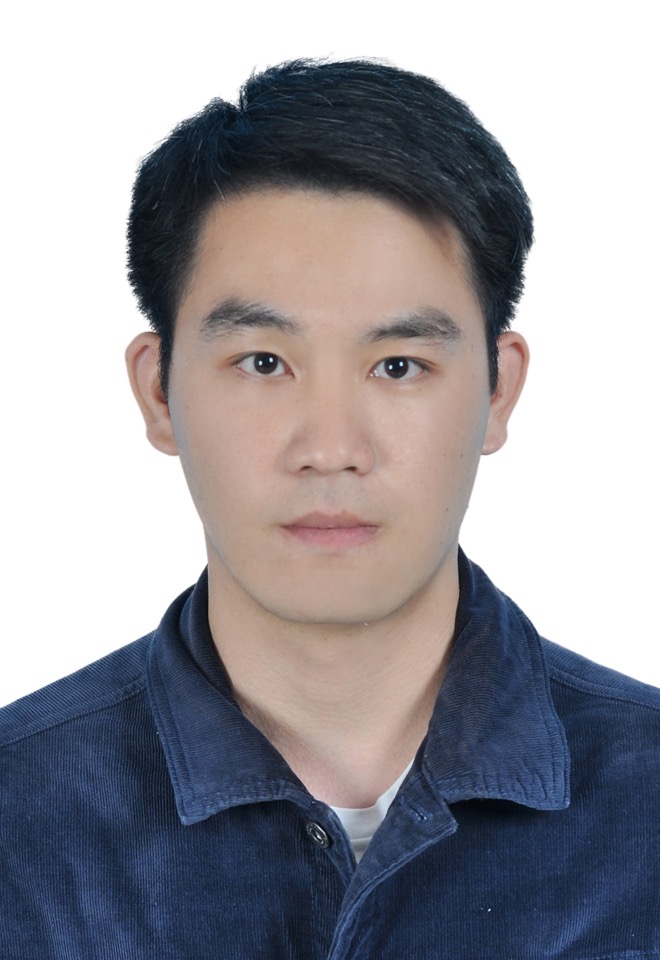}}]{Guoji Fu} received the B.Eng. degree from the University of Science and Technology of Beijing in 2017,
and the MSc degree from the Joint Master's Program between Southern University of Science and Technologys, Shenzhen, China and Harbin Institute of Technology, Harbin, China in 2019.
His research interests include machine learning and data mining.
\end{IEEEbiography}

\begin{IEEEbiography}[{\includegraphics[width=1in,height=1.25in,clip,keepaspectratio]{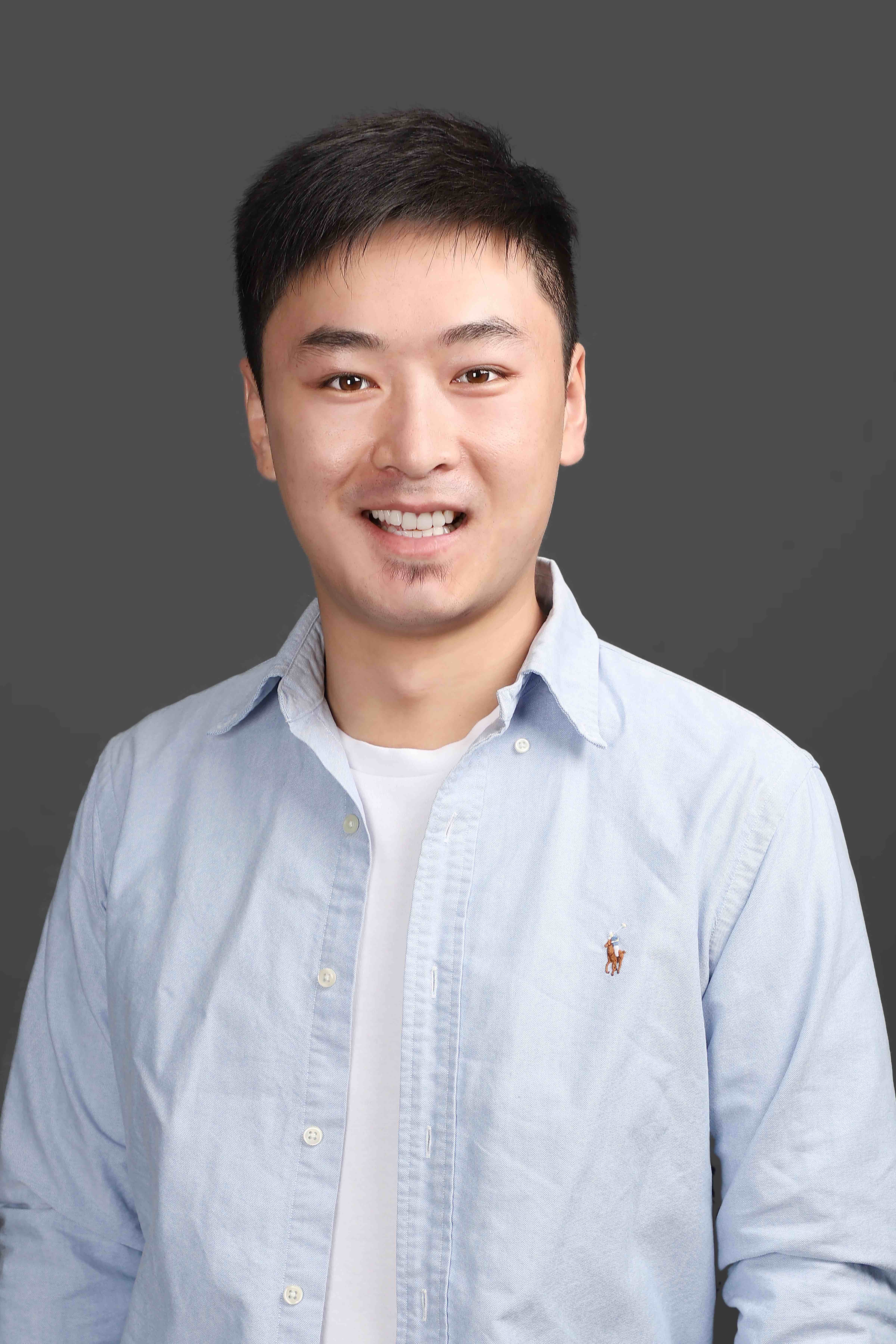}}]{Peng Yang} (S'14-M'18) received the B.Eng. degree and Ph.D degree in computer science and technology from the University of Science and Technology of China (USTC), Hefei, China, in 2012 and 2017, respectively. From 2018, He has been working as a Research Assistant Professor at the Department of Computer Science and Engineering, Southern University of Science and Technology, Shenzhen, China. His research interests include Evolutionary Policy Optimization and its applications. He has been served as a regular reviewer of several world-class journals and a program committee member of a set of top international conferences.
  \end{IEEEbiography}

\begin{IEEEbiography}[{\includegraphics[width=1in,height=1.25in,clip,keepaspectratio]{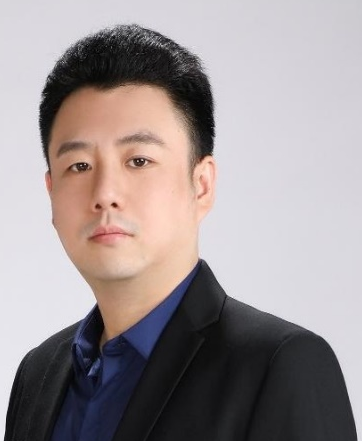}}]{Zheng Hu} Dr. Zheng Hu is the Director of Reliability Advanced Technology Lab, Huawei, a leading global provider of information and communications technology (ICT) infrastructure and smart devices. He is currently leading a corporate-level project, Trustworthy AI, in charge of the research and innovation of key technologies towards the reliable and safe AI system. Meanwhile his research also focuses on the software reliability, reliability theory and ah-hoc networks, etc. He received his PhD degree in Computer Science from Lyon University in Lyon France. Before joint Huawei, he was the senior researcher in Orange Labs (France Telecom), working on the self-configuration network of smart home/smart building.
\end{IEEEbiography}

\begin{IEEEbiography}[{\includegraphics[width=1in,height=1.25in,clip,keepaspectratio]{Author4}}]{Shan He} received the Ph.D. degree in electrical engineering and electronics from University of Liverpool, Liverpool, U.K., in 2007. He is a Senior Lecturer (Tenured Associate Professor) in School of Computer Science, the University of Birmingham. He is also an affiliate of the Centre for Computational Biology. His research interests include complex networks, machine learning, optimisation and their applications to medicine. He is an Associate Editor of IEEE Transactions on Nanobioscience.
\end{IEEEbiography}

\begin{IEEEbiography}[{\includegraphics[width=1in,height=1.25in,clip,keepaspectratio]{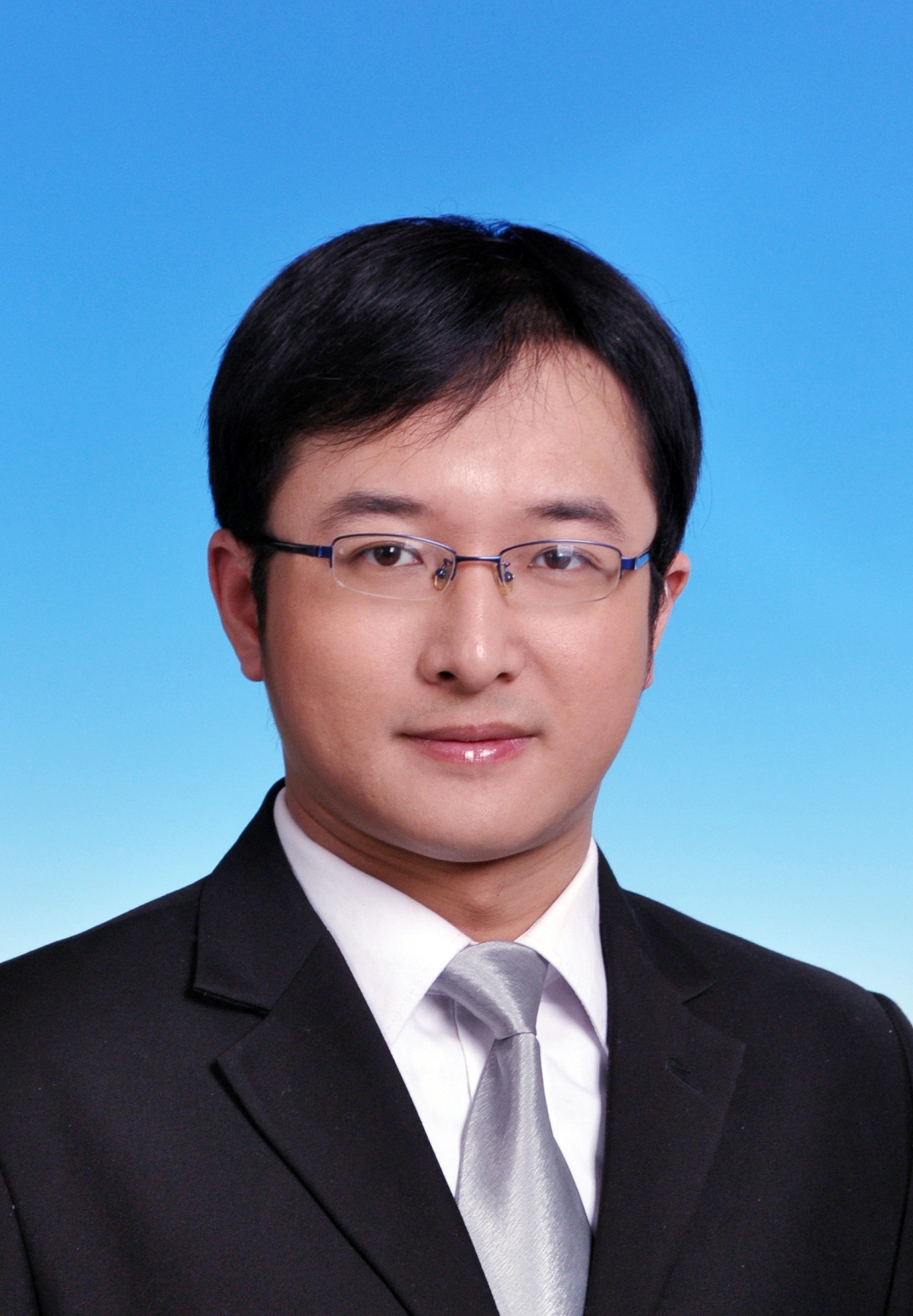}}]{Ke Tang} (Senior Member, IEEE) received the B.Eng. degree from the Huazhong University of Science and Technology, Wuhan, China, in 2002 and the Ph.D. degree from Nanyang Technological University, Singapore, in 2007. From 2007 to 2017, he was with the School of Computer Science and Technology, University of Science and Technology of China, Hefei, China, first as an Associate Professor from 2007 to 2011 and later as a Professor from 2011 to 2017. He is currently a Professor with the Department of Computer Science and Engineering, Southern University of Science and Technology, Shenzhen, China. He has published more than 180 papers, which have received over 10,000 Google Scholar citations with an H-index of 50. His current research interests include evolutionary computation, machine learning, and their applications.  

Prof. Tang received the IEEE Computational
Intelligence Society Outstanding Early Career Award (2018), the Newton
Advanced Fellowship (Royal Society, 2015) and the Natural Science Award
of Ministry of Education of China (2011 and 2017), and is a Changjiang
Scholar Professor (awarded by the MOE of China).
He is an Associate Editor of the IEEE Transactions on Evolutionary Computation and served as a member of Editorial Boards for a few other journals.
\end{IEEEbiography}

\end{document}